\documentclass{aa}  

\usepackage{graphicx}
\usepackage{txfonts}
\usepackage{lipsum}
\usepackage{subcaption}         
\usepackage{lscape}
\usepackage{placeins}           
\usepackage{orcidlink}
\usepackage{hyperref}

\begin{document}

\title{Misalignment of the Lense-Thirring precession by an accretion torque}

\author{
    D.A. Bollimpalli 
    \orcidlink{0000-0001-8835-8733}
    \inst{1,2}
    \and
    J. Hor\'{a}k
    \orcidlink{0000-0002-7635-4839}
    \inst{3,4}
    \and
    W. Kluźniak
    \orcidlink{0000-0001-9043-8062}
    \inst{4}
    \and
    P. C. Fragile
    \orcidlink{0000-0002-5786-186X}
    \inst{5,6}
}

\institute{
    Department of Astronomy, Astrophysics \& Space Engineering, Indian Institute of Technology Indore, Simrol, Indore 453552, Madhya Pradesh, India \\
    \email{dbollimpalli@iiti.ac.in}
    \and
    Center for Interdisciplinary Exploration \& Research in Astrophysics (CIERA), Physics \& Astronomy, Northwestern University, Evanston, IL 60202, USA
    \and
    Astronomical Institute, Academy of Sciences, Bo\v{c}n\'{\i}~II 141\,31 Prague~4, Czech Republic\\
    \email{horak@astro.cas.cz}
    \and
    Nicolaus Copernicus Astronomical Center, ul. Bartycka 18, PL 00-716 Warsaw, Poland\\
    \email{wlodek@camk.edu.pl}
    \and
    Department of Physics and Astronomy, College of Charleston, 66 George St, Charleston, SC 29424, USA \\
    \and
    Center for Computational Astrophysics, Flatiron Institute, 162 5th Avenue, New York, NY 10010, USA
}

\date{Received , 2025; accepted , 2025}

\abstract
{Orbiting matter misaligned with a spinning black hole undergoes Lense-Thirring precession, due to the frame-dragging effect. This phenomenon is particularly relevant for type-C QPOs observed in the hard states of low-mass X-ray binaries. However, the accretion flow in these hard states is complex, consisting of a geometrically thick, hot corona surrounded by a geometrically thin, cold disk. Recent simulations have demonstrated that, in such a truncated disk scenario, the precession of the inner hot corona slows due to its interaction with the outer cold disk.}
{This paper aims to provide an analytical description of the precession of an inner (hot) torus in the presence of accretion torques exerted by the outer (cold) disk.}
{Using the angular momentum conservation equation, we investigate the evolution of the torus angular momentum vector for various models of accretion torque.}
{We find that, in general, an accretion torque tilts the axis of precession away from the black hole spin axis. In all models, if the accretion torque is sufficiently strong, it can halt the precession; any perturbation from this stalled state will cause the torus to precess around an axis that is misaligned with the black hole spin axis.}
{The accretion torque exerted by the outer thin disk can cause precession around an axis that is neither aligned with the black hole spin axis nor perpendicular to the plane of the disk. This finding may have significant observational implications, as the jet direction, if aligned with the angular momentum axis of the torus, may no longer reliably indicate the black hole spin axis or the orientation of the outer accretion disk.}


\maketitle
\nolinenumbers
%-------------------------------------------------------------

\section{Introduction}
Quasi-periodic oscillations (QPOs) in the X-ray light curves of accreting black hole and neutron star X-ray binary systems, characterized by broad peaks in their power spectra, are a characteristic feature of these systems \citep[see][for reviews]{VanDerKlis2006, Remillard+McClintock2006}. The "quasi-periodic" nature arises from the modulation of either the frequency or the amplitude of the light curves. QPOs offer a unique opportunity to study the strong gravitational fields surrounding compact objects, providing insights into physical processes on spatial scales that are otherwise inaccessible, with the exception of nearby supermassive black holes observed by the Event Horizon Telescope. Based on their frequency, black hole QPOs are typically categorized as either high frequency ($\geq$ 60 Hz) or low ($\leq 30$ Hz) \citep{Psaltis+Belloni+VanDerKlis1999, Ingram+Done2011}.

In addition to rapid variability, accreting black hole X-ray binaries exhibit spectral state transitions over periods ranging from days to months \citep[e.g.][]{Homan+Belloni2005, BlackCAT, Singh+2019}. These transitions, often depicted as q-shaped tracks in hardness-intensity diagrams, reflect significant changes in the geometry and physical conditions of the accretion flow \citep{FBG2004, Belloni+2005, MR2006, Weng+2021}. The primary spectral states are classified as "soft" and "hard," though intermediate states have also been observed \citep{Belloni2010, Done+2007}. In the soft state, the X-ray spectrum is dominated by thermal emission, resembling blackbody radiation from a cool, optically thick, geometrically thin accretion disk \citep{Shakura+Sunyaev1973}. In contrast, the hard state is dominated by a non-thermal, power-law emission produced by Compton up-scattering of soft seed photons from the disk by a hot, optically thin electron cloud, commonly referred to as the "corona" \citep{Haardt+Maraschi1991, Haardt+Maraschi1993, Zdziarski+1998, Zdziarski+2004, Fabian+2000, Krawczynski+Beheshtipour2022, Groselj+2024}.

While the precise geometry of the corona remains uncertain, a widely accepted model envisions a truncated disk, where the inner thin disk is replaced by a corona \citep{Esin+McClintock+Narayan1997, Done+2007}.
As the system evolves from the hard to the soft state, the disk migrates inward, and the corona collapses, eventually reaching the innermost stable circular orbit \citep{Plant+2015, Kara+2019, Buisson+2019, DeMarco+2021, Rawat+2025}.

QPOs are observed across different spectral states, with low-frequency QPOs (LFQPOs) being particularly prominent \citep{Wijnands+1999, Casella+Belloni+Stella2005}. Type-A and type-B LFQPOs typically appear during transitions through the intermediate states toward the soft state, while type-C LFQPOs appear mostly in the hard state. High-frequency QPOs, on the other hand, tend to occur in high-luminosity accretion states \citep{Remillard+2002}. Several models attribute QPOs to oscillations in the corona or the thin disk, or to waves generated by instabilities within the accretion flow\footnote{For a comprehensive review of QPOs and the suggested models, we refer the readers to \citet{IM2019} and references therein.}. Understanding the origin and mechanisms driving QPOs may be helpful in constraining the geometry and location of the corona.

One of the most common interpretations of the type-C LFQPO is that it results from Lense-Thirring precession of the corona. This hypothesis is supported by observations showing that the QPO amplitude and phase lag between hard and soft photons, measured at the QPO frequency, depend strongly on the inclination angle of the system \citep{Motta2015, eijnden2017}. In the truncated-disk model, the precession of the corona, surrounded by the truncated disk, produces the type-C LFQPOs \citep{Ingram+Done+Fragile2009, Ingram+Done2011}. This model also explains the observed correlation between the low-frequency break (attributed to the viscous timescale at the transition radius) and the type-C QPO frequency (attributed to the precession frequency of the inner hot flow) in the power spectra of accreting black hole systems. The model has been further explored in great detail to include time-dependent emission of iron-line profiles arising from the variable orientation of the hot inner flow and the outer thin disk with respect to the observer \citep{Ingram+Done2012b, Ingram+2016}, correlation of the X-ray \citep{Zycki+2016} and optical \citep{Veledina+Poutanen2015} spectral properties with QPO phase, as well as variable X-ray polarization \citep{Ingram+2015, Fragile25}.  

 A sufficiently hot, orbiting fluid around a Kerr black hole may take the shape of a torus. Among the eigenmodes of this torus (assuming it is slender) are rigid-body motions, the simplest being a uniform vertical\footnote{The vertical direction is taken to be along the spin axis of the black hole.} oscillation at a frequency corresponding to the vertical epicyclic frequency of a test particle, $\sigma=\omega_{\perp}$ \citep{Bursa+2004}. This eigenmode is often invoked to explain high-frequency QPOs observed in X-ray binaries. This mode, characterized by identical vertical displacements of all fluid elements, corresponds to the $m=0$ mode in a more general class of motions where vertical displacement varies azimuthally as  $\exp(i m\phi)$. The pattern frequency of these oscillations\footnote{For a discussion of general modes (involving internal oscillations) of relativistic slender tori, see \citet{Blaes+2006}. We take the displacement to be proportional to $\exp[\mathrm{i}(-\omega t+ m\phi)]$.} is given by $\omega_\mathrm{p}=(\sigma + m\Omega)/m$, where $\Omega$ is the orbital frequency. The $m=-1$ mode, in particular, corresponds to a rigid-body precession of an inclined torus at a frequency defined by the difference between the orbital and vertical epicyclic frequencies:
\begin{equation}
    \omega_\mathrm{p}=\Omega - \omega_\perp.   
    \label{eq1}
\end{equation}
This precession mechanism can be understood by considering the orbit of a test particle inclined at an angle to the equatorial plane of the black hole. Due to the frame-dragging effect of the spinning black hole, the orbit undergoes nodal precession, known as Lense-Thirring precession \citep{LT1918}. The  vertical, epicyclic, rigid-body mode is thus thought to explain both the high and low-frequency QPOs (for $m=0$ and $m=-1$, respectively) observed in X-ray binaries. \citep[A similar combination of the {\sl radial} epicyclic and orbital frequencies was suggested as an explanation for kHz QPOs in neutron stars by][]{SV98, SVM99}.

A torus with a substantial radial extent can also undergo nearly rigid-body precession if the sound travel time across the torus is shorter than the precession period, $2\pi/\omega_\mathrm{p}$. In this case, the Lense-Thirring torque is efficiently communicated through bending waves, causing the torus to behave as a solid body \citep{Papaloizou+Terquem1995, Lodado+Facchini2013, Nixon+King2016}. The precession periods of non-slender tori have been derived analytically using perturbative expansions of the fluid equations \citep{Blaes+2007} \citep[see also][for a full general relativistic treatment]{Straub+Sramkova2009}, and these precession frequencies match the form of equation~(\ref{eq1}), with radial averages replacing the orbital and vertical frequencies \citep{Blaes+2007}.

Numerous general relativistic magneto-hydrodynamic (GRMHD) simulations have confirmed the robustness of such nearly rigid-body Lense-Thirring precession of accreting tori \citep{Fragile+Anninos2005, Fragile+2007, Teixeira+2014, Liska+2018}. In particular, \citet{Fragile+2007} showed that the numerically derived precession rates agree well with the precession frequency for a rigid body \citep{Liu+Melia2002}. Their derived expression for the precession frequency is a function of the inner and outer radii of the torus, based on the density and rotation profiles extracted from their simulations. These formulas have since been refined by \citet{Ingram+Done+Fragile2009, Ingram+Done2011, Ingram+Done2012a} and \citet{DeFalco+Motta2018}.

Note that in this case of free Lense-Thirring precession, the `tilt' angle between the axes of the torus and the black hole remains constant over time. This means the angular momentum vector of the torus rotates uniformly about the black hole  spin axis, sweeping out a conical surface.

However, in an astronomical context, the torus is typically the inner region of an accretion flow, surrounded by a disk. Notably, none of the studies listed above account for the effect of the outer disk on the torus precession. Only recently have GRMHD simulations of tilted, truncated accretion disks with inner tori demonstrated that the exchange of angular momentum between the outer thin disk and the torus introduces an accretion torque, which significantly alters the torus  precession dynamics \citep{Bolimpalli+2023}. 

In this paper, we examine analytically the leading effects of an accretion torque on the Lense-Thirring precession of an inner torus within this theoretical context. We find that the torus precesses around an axis different from the black hole spin axis. In other words, the precession occurs around an axis that is neither aligned with the black hole spin axis, nor perpendicular to the plane of the accretion disk. This finding has important observational implications: if the jet is anchored in the torus and shares its mean angular-momentum axis, the jet itself need not be aligned with the black hole spin. Recent IXPE observations showing X-ray polarization roughly aligned with the jet direction \citep{Krawczynski+Beheshtipour2022, Zdziarski+2023} suggest a common corona–jet axis, but this axis does not necessarily coincide with the spin. This scenario favors jet formation mechanisms influenced by the accretion flow \citep[e.g.,][or hybrid models]{Blandford+Payne1982}, while a pure \citet{Blandford+Znajek1977} jet anchored strictly to the horizon would instead align with the spin. Unfortunately, rapid precession of the torus or small-scale jet wobbling could make these effects difficult to detect observationally. Additionally, magnetic topology or GRMHD effects could further modulate the effective jet axis.

The rest of the paper is organized as follows: In \S~\ref{sec2}, we present the basic framework of our model, which is based on the truncated disk geometry. \S~\ref{sec3} derives the general solution for  the precession of a tilted, accreting torus. In \S~\ref{sec:solutions}, we provide specific examples of different accretion torques and demonstrate how precession occurs in each case. \S~\ref{sec5} compares our results with recent simulation findings from \citet{Bolimpalli+2023}. Finally, in \S~\ref{sec:discus}, we discuss the observational implications and present our conclusions.

%-------------------------------------------------------------

\section{The overall picture}
\label{sec2}
\begin{figure}[ht!]
    \centering
    \includegraphics[width=\hsize]{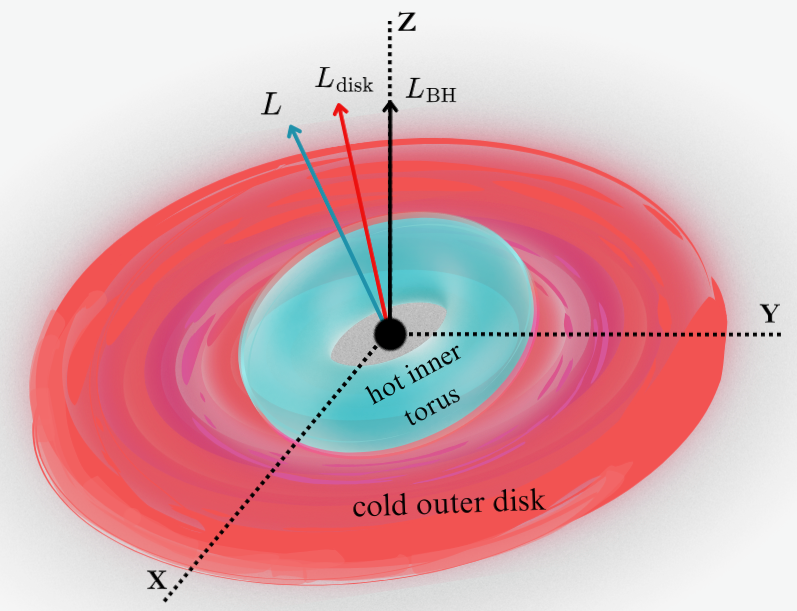}
    \caption{Geometry of our model. The direction of the black-hole angular momentum $\boldsymbol{L}_\mathrm{BH}$ coincides with the $z$-direction of the Cartesian system used in the calculations. Due to misalignment with the black-hole spin, the angular momenta of the outer  thin accretion disk ($\boldsymbol{L}_\mathrm{disk}$) and inner hot torus ($\boldsymbol{L}$) have nonzero projections into the $x$--$y$ plane.}
    \label{fig:sketch}
\end{figure}

We imagine a spinning black hole with its axis misaligned with respect to the accretion flow. We consider the two-component accretion flow geometry that is thought to be relevant for hard states of low-mass X-ray binaries (LMXBs), i.e., an inner, hot torus surrounded by an outer cool, thin disk, as shown in Fig.~\ref{fig:sketch}. Since the inner torus is geometrically thick and optically thin, bending waves are communicated efficiently enough for the torus to precess as a rigid-body\footnote{If the inner torus were not accreting from the outer disk, it would be able to undergo free Lense-Thirring precession, as discussed in the introduction.}. On the other hand, due to the geometrically thin and optically thick nature of the outer disk, the bending waves are diffused away there and we expect no precession in this region.  

The accretion flow is tilted in that if we align the $z$ axis of a Cartesian coordinate frame with the spin axis of the black hole (the "vertical" direction, see Fig.~\ref{fig:sketch}), the angular momentum vector of the disk has a nonzero projection onto the $x$-$y$ plane. Thus, any accretion from the disk onto the torus will tend to change the $x$-$y$ component of the torus angular momentum through the advection of the disk  angular momentum (in addition to affecting the $z$ component). This will result in a rotation in the $x$-$y$ plane of the projection of the angular momentum onto it, beyond any relativistic precession, if the torus is already tilted with respect to the black hole spin axis. Alternatively, it will tilt the torus if it happens to be aligned with the black hole. The net advection rate of the angular momentum into the torus is equivalent to a torque, which we call the accretion torque. Our aim is to derive the time evolution of the angular momentum of the torus for a given accretion torque, and in particular to discuss the direction in which the torus axis will point. More specifically, we are interested in the evolution of the precession angle of the torus, defined as the angle from the $x$ axis to the $x$-$y$ projection of the torus angular momentum vector.

%-------------------------------------------------------------

\section{The torque equation}
\label{sec3}
For our purposes, it is sufficient to use a Newtonian equation of angular momentum conservation,
\begin{equation}
    \frac{\mathrm{d}\boldsymbol{L}}{\mathrm{d}t}=\boldsymbol{\tau},
	\label{eq:inh}
\end{equation}
with
$\boldsymbol{\tau}=\boldsymbol{\tau}_\mathrm{p} + \boldsymbol{\tau}_\mathrm{acc}$.
We include the main effect of general relativity by adopting a Lense-Thirring torque
\begin{equation}
    \boldsymbol{\tau}_\mathrm{p} = \boldsymbol{\omega}_\mathrm{p}\times \boldsymbol{L},
\end{equation}
where $\boldsymbol{\omega}_\mathrm{p} = 2(G/c^2)(\boldsymbol{L}_\mathrm{BH}/R^3)$ is a constant vector in the vertical (black hole spin) direction, $\boldsymbol{L}_\mathrm{BH}$ is the black-hole angular momentum, and $R$ is a torus mean radius \citep{Wilkins1972, Bardeen+Patterson1975}. 
In effect, we assume that the torus has a constant free precession rate, $\boldsymbol{\omega}_\mathrm{p}$, regardless of its mass and total angular momentum $\boldsymbol{L}$. The second component of the torque, $\boldsymbol{\tau}_\mathrm{acc}$, is the accretion torque equal to the net rate of angular momentum advection. Thus, the evolution of the torus angular momentum is given by
\begin{equation}
    \frac{\mathrm{d}\boldsymbol{L}}{\mathrm{d}t}= \boldsymbol{\omega}_\mathrm{p} \times \boldsymbol{L} + \boldsymbol{\tau}_\mathrm{acc}.
	\label{eq:inho}
\end{equation}
Note that we have ignored the viscous terms in the above conservation equation, as they were found to be negligible when compared to the Lense-Thirring torque and angular momentum fluxes in the simulations of \citet{Bolimpalli+2023}. 

In the Cartesian coordinate system introduced in the previous Section, $\boldsymbol{\omega}_\mathrm{p}=\omega_\mathrm{p} \boldsymbol{\hat{z}}$, and hence $\boldsymbol\tau_\mathrm{p}$ does not depend on $L_z$. Therefore the $z$ component of the torque equation 
\begin{equation}
    \frac{\mathrm{d}L_z}{\mathrm{d}t}= (\boldsymbol{\tau}_\mathrm{acc})_z
	\label{eq:lz}
\end{equation}
does not couple to the $x$ and $y$ components, and it is the latter components that we focus on in our analysis. Introducing complex variables: 
\begin{equation}
    L\equiv L_x +\mathrm{i}L_y,
    \label{eq:Lcomp}
\end{equation}
so that $L_x=\mathrm{Re}\,L$, $L_y=\mathrm{Im}\,L$,  and similarly for $\boldsymbol\tau_\mathrm{acc}$,
we can rewrite the $x$ and $y$ components of Eq.~(\ref{eq:inho}) as
\begin{equation}
    \frac{\mathrm{d}L}{\mathrm{d}t}=\mathrm{i}\omega_{\rm p} L + \tau_\mathrm{acc}.
	\label{eq:inhomo}
\end{equation}
For $\tau_\mathrm{acc}=0$ the equation has a simple solution corresponding to uniform rotation in the positive direction
\begin{equation}
    L(t)=L_0\exp(\mathrm{i}\omega_{\rm p} t)\equiv L_\mathrm{LT}(t).
	\label{eq:lt}
\end{equation}
This is just the $x$-$y$ projection of the angular momentum vector of the torus undergoing Lense-Thirring free precession, hence the notation.

Note that precession may be halted for a sufficiently large accretion torque (or sufficiently small tilt angle of the torus). It suffices that the following equation be satisfied, starting at some instant $t_0$,
\begin{equation}
    \boldsymbol{\tau}_\mathrm{acc}= - \boldsymbol{\omega}_{\rm p} \times \boldsymbol{L},
\end{equation}
for $d\boldsymbol{L}/dt=0$ to hold for $t>t_0$. In the $x$-$y$ plane, this is just
\begin{equation}
   \tau_\mathrm{acc} = -\mathrm{i}\omega_{\rm p} L,
	\label{eq:librium}
\end{equation}
and has a simple interpretation. The Lense-Thirring torque being perpendicular to the angular momentum vector leads it in phase by $\mathrm{\pi/2}$ (factor of $\mathrm{i}$), so an accretion torque of equal magnitude but in opposite direction will counteract it, also being perpendicular to the angular momentum vector, but lagging it by the same phase angle $\mathrm{\pi/2}$ (factor of $-\mathrm{i}$). 

However, this equilibrium point, let us call it $\boldsymbol{L}_1$, corresponding to $\boldsymbol{\tau}_\mathrm{acc}=\boldsymbol{\tau}_1$ is not stable: 
a small fluctuation of $\boldsymbol{L}$ will lead to resumption of precession (about the equilibrium point $\boldsymbol{L}_1$). Indeed, keeping $\tau_1= -\mathrm{i}\omega_{\rm p} L_1$ fixed, a change of $L$ from $L_1$ to $L_2$ at time $t_2$ will lead to the resumption of precession for $t>t_2$ with $L_0=L_2-L_1$, by Eqs.~\ref{eq:inhomo}, \ref{eq:lt}, \ref{eq:librium}.
A similar result would be obtained upon an impulsive change of the accretion torque
$ \tau_\mathrm{acc}$ from the value $\tau_1$ to $\tau_2$, with the precession now occurring around the point of equilibrium $L_2=\mathrm{i}\tau_2/\omega_{\rm p}$ with amplitude $L_0=L_1-L_2$. This is a generic result: a change of the accretion torque leads to a change in the precession, and in particular can induce a Lense-Thirring precession where there was none before, as will also be seen in our discussion in Subsection~\ref{healing}.

The general solution of the inhomogeneous linear differential Eq.~(\ref{eq:inho}) is the sum of the general solution of the homogeneous equation
\begin{equation}
    \frac{\mathrm{d}L}{\mathrm{d}t}=\mathrm{i}\omega_{\rm p} L,
	\label{eq:homo}
\end{equation}
which we already know to be $L_\mathrm{LT}(t)$, and a special solution of the inhomogeneous Eq.~(\ref{eq:inho}). Thus,
\begin{equation}
    L(t)=L_\mathrm{LT}(t) + \hat{L}(t),
	\label{eq:general}
\end{equation}
where $\hat{L}$ is a particular solution satisfying 
\begin{equation}
    \frac{\mathrm{d}\hat{L}}{\mathrm{d}t}=\mathrm{i}\omega_{\rm p} \hat{L} + 
    \tau_\mathrm{acc}.
	\label{eq:inhomage}
\end{equation}

To summarize, the general solution to the motion of a tilted, accreting torus is a precession (Eq.~[\ref{eq:lt}]) at the Lense-Thirring frequency around a (possibly moving) axis $\hat{L} \equiv \hat{L}_x + i\hat{L}_y$, given by the solution of Eq.~(\ref{eq:inhomage}), which can be taken as:
\begin{equation}
    L(t) = \left( L_0 + \int \tau_{\rm acc}(t^\prime) \exp{(-i\omega_{\rm p}t^\prime)}{\rm d} t^\prime \right) \exp{(i\omega_{\rm p} t)}.
\end{equation}
All that remains is to find $\hat{L}(t)$ and $L(t)$ for any given $\tau_\mathrm{acc}(t)$, which is what we will address in the next section.

%-------------------------------------------------------------

\section{Solutions}
\label{sec:solutions}
We now give solutions of Eq.~(\ref{eq:inhomage}) for four simple forms of $\tau_\mathrm{acc}$. In steady state, it may be reasonable to assume that the net rate of angular momentum advection is constant in time. This corresponds to a steady torque (\S~\ref{steady}). On the other hand, the luminosity of X-ray binaries often varies in time. A steady rise in advected angular momentum rate (\S~\ref{rise}) may correspond to black hole disks going into ``outburst.'' A change of the accretion torque from one value to another (\S~\ref{healing}) could perhaps model state transitions of black hole binaries, or perhaps the torque could oscillate (\S~\ref{oscillating}). Finally, it is possible that the ring of the accretion disk directly adjacent to the torus may itself precess, for instance in the corrugation mode of diskoseismology \citep{Kato1989, Silbergleit2001, Kato1993, Ipser1996}. The $x$-$y$ components of the advected angular momentum could then vary harmonically, so that the accretion torque could be taken to rotate at a definite frequency (\S~\ref{rotating}).

\subsection{Steady accretion torque}
\label{steady}
\begin{figure}[ht!]
    \centering
    \includegraphics[width=\hsize]{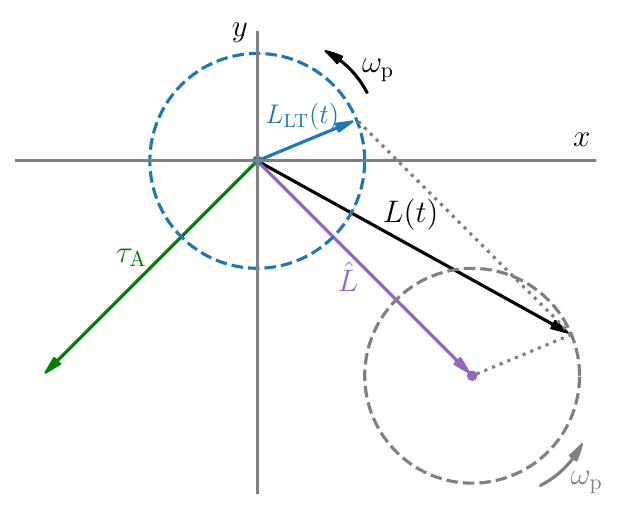}
    \caption{Precession of the angular momentum vector $L$ of the inner torus when a steady accretion torque $\tau_\mathrm{f}=\tau_\mathrm{A}$ is applied. This figure is projected into the plane perpendicular to the black-hole spin. In the absence of an accretion torque, the angular momentum vector executes free Lense-Thirring precession $L_\mathrm{LT}(t)$ around the direction of the black hole spin. When the torque is applied, the general solution consists of free Lense-Thirring precession about a new direction (shifted by a constant, $\hat{L}$).}
    \label{fig:steady-torque}
\end{figure}

For a steady torque, $\tau_\mathrm{acc}=\tau_\mathrm{f}=\mathrm{const}$, the solution is
\begin{equation}
    \hat{L} = \frac{\mathrm{i}\tau_\mathrm{acc}}{\omega_{\rm p}}=
    \frac{\mathrm{i}\tau_\mathrm{f}}{\omega_{\rm p}}=\mathrm{const}.
    \label{eq:steady}
\end{equation} 
The general solution of Eq.~(\ref{eq:inhomo}) is then 
\begin{equation}
    L(t) = \frac{\mathrm{i}\tau_\mathrm{f}}{\omega_{\rm p}} + 
    L_0\exp\left(\mathrm{i}\omega_{\rm p} t\right),
    \label{eq:steadygeneral}
\end{equation}
describing a precession with the Lense-Thirring frequency about a point (in the $x$-$y$ plane) which is advanced by $\mathrm{\pi}/2$ in phase with respect to the accretion torque (see Fig.~\ref{fig:steady-torque}). The reason for this has already been given in the paragraph following Eq.~(\ref{eq:librium}). The amplitude of precession, $L_0$, is given by the initial conditions.

\subsection{Linearly growing torque}
\label{rise}
Taking $\tau_\mathrm{acc}(t)=\tau_\mathrm{0}+ (t/t_\mathrm{f})\tau_\mathrm{f}$, we obtain
\begin{equation}
    \hat{L}(t)=\frac{\tau_\mathrm{f}}{t_\mathrm{f}\omega_{\rm p}^2} + 
    \frac{\mathrm{i}\tau_\mathrm{acc}(t)}{\omega_{\rm p}}.
    \label{eq:rise}
\end{equation}
In this case, the $x$-$y$ projection of the tip of the angular momentum vector undergoes uniform circular motion at the Lense-Thirring precession rate, similar to the case in section~\ref{steady}. However, in this scenario, the point around which this precession occurs will experience linear drift in the direction that advances the accretion torque by $\pi/2$ in phase. In the limit of slow growth, $\omega_{\rm p} t_\mathrm{f}\gg 1$, this tends to the steady torque solution, albeit with a varying position of the center of precession, $\hat{L}(t)\approx\mathrm{i} \tau_\mathrm{acc}(t)/\omega_{\rm p}$.

\begin{figure*}[ht!]
    \centering
    \includegraphics[width=\hsize]{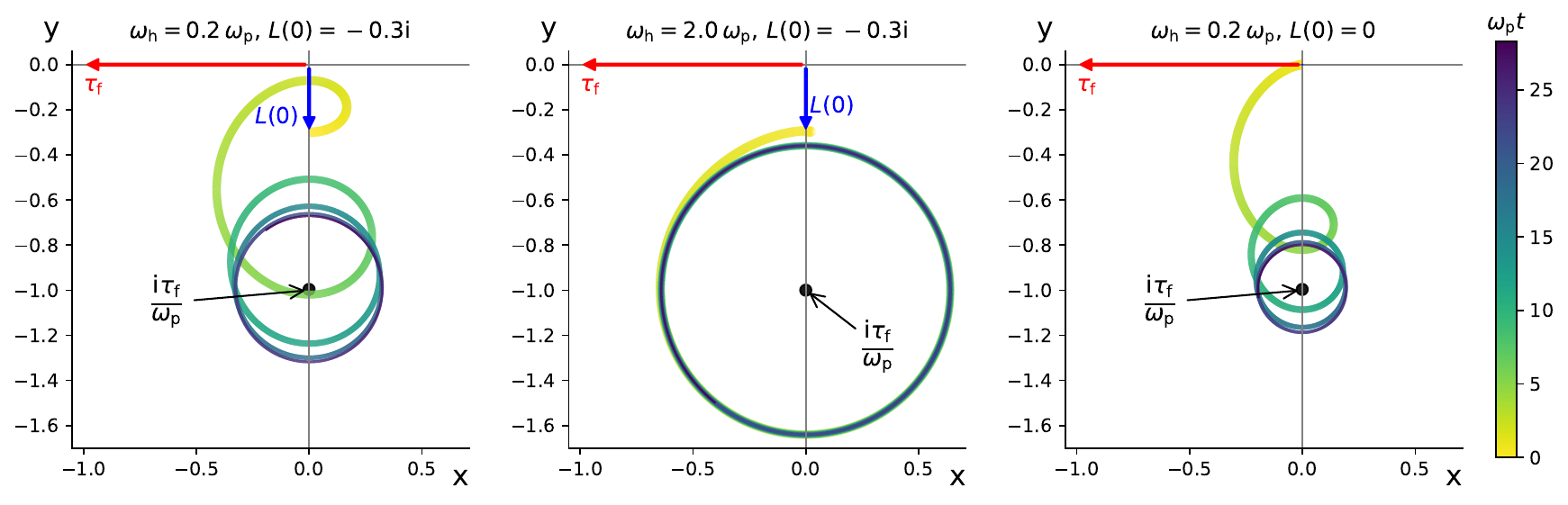}
        \caption{Trajectory of the torus angular-momentum vector in the $x$--$y$ plane when the accretion torque evolves exponentially from zero to $\tau_\mathrm{f} = -1$ for various cases. A torus that is initially precessing with an angular momentum $L(0) = -0.3{\rm i}$ approaches the final state of precession about the new equilibrium point, $\mathrm{i}\tau_\mathrm{f}/\omega_{\rm p}$, slowly or rapidly for a low ($\omega_\mathrm{h} = 0.2\omega_\mathrm{p}$) or high ($\omega_\mathrm{h} = 2\omega_\mathrm{p}$) healing frequency, as shown in the {\sl left} and {\sl middle} panels, respectively. The  {\sl right} panel presents the same case as in the {\sl left} (i.e., $\omega_\mathrm{h} = 0.2\omega_\mathrm{p}$), except that initially the torus has a zero angular momentum projection onto the $x$--$y$ plane, i.e., $L(0) = 0$, highlighting our finding that the change in accretion torque can induce precession in a torus. In all three examples, we took $\omega_\mathrm{p} = 1$ and the phase evolution is represented by both the colour and the decreasing thickness of the curve.}
    \label{fig:healing-trajectory}
\end{figure*}

\subsection{Change of torque}
\label{healing}
Let us consider a variation in torque that is exponentially changing from $\tau_\mathrm{i}$ to $\tau_\mathrm{f}$, with a decay constant $\omega_\mathrm{h}$,
\begin{equation}
    \tau_\mathrm{acc}(t)={\tau_\mathrm{f}} + 
    \left({\tau_\mathrm{i}}-{\tau_\mathrm{f}}\right)
    \exp\left(-\omega_\mathrm{h}t\right).
\end{equation}
In this case, the particular solution takes the form
\begin{equation}
    \begin{aligned}
        \hat{L}(t) = &
        -\frac{\omega_\mathrm{h}}{\omega_{\rm p}^2+\omega_\mathrm{h}^2}\left({{\tau_\mathrm{i}}-\tau_\mathrm{f}}\right)\exp(-\omega_\mathrm{h}t)
        \\
        & \quad +\,\mathrm{i}\left[\frac{\tau_\mathrm{f}}{\omega_{\rm p}}\,+ \frac{\omega_{\rm p}}{\omega_{\rm p}^2+\omega_\mathrm{h}^2}\left({\tau_\mathrm{i}}-{\tau_\mathrm{f}}\right)\exp(-\omega_\mathrm{h}t)\right].
        \label{eq:healing}
    \end{aligned}
\end{equation}
This tends exponentially to the constant torque solution. When the change is slow, $\omega_\mathrm{h}\ll\omega_{\rm p}$, this approximates the steady torque solution, with a different value of the accretion torque at every instant of time: $\hat{L}(t)\approx\mathrm{i} \tau_\mathrm{acc}(t)/\omega_{\rm p}$.
In the opposite limit of fast healing, $\omega_{\rm p}\ll\omega_\mathrm{h}$, the solution is very close to the steady torque solution with the final value of the torque, $\hat{L}(t)\approx\mathrm{i} \tau_\mathrm{f}/\omega_{\rm p}$. 

A general solution consists of a free Lense-Thirring precession (\ref{eq:lt}) and the special solution (\ref{eq:healing}), whose properties were discussed above. Fig.~\ref{fig:healing-trajectory} shows two such examples of how the torus angular momentum evolves for the slow ($\omega_{\rm h} = 0.2 \omega_{\rm p}$; left panel) and fast ($\omega_{\rm h} = 2 \omega_{\rm p}$; middle panel) change of the accretion torque from 0 to $\tau_{\rm f}$. Since the accretion torque is initially zero, the torus precesses around the $z$-axis. Interestingly, even if the initial torus is aligned with the black hole spin-axis, i.e., precession is zero, the accretion torque change can still excite free precession of the torus, as shown in the last panel of Fig~\ref{fig:healing-trajectory}. 

\subsection{Oscillating torque}
\label{oscillating}
It turns out that none of the three simple torque models in Sections~\ref{steady}, \ref{rise}, \ref{healing}, provide a good description of the simulations of \citet{Bollimpalli+2024}. One could similarly consider a change in torque that is oscillating between $\tau_{\rm A}$ and $\tau_{\rm B}$ at a frequency $\omega_{\rm h}$:
\begin{eqnarray}
    \tau_{\rm acc}(t) =\frac{\left({{\tau_\mathrm{A}}+\tau_\mathrm{B}}\right)}{2} +\frac{(\tau_{\rm B} - \tau_{\rm A})}{2} \sin(\omega_{\rm h}t), \nonumber\\
    \label{eq:osilltorq}
\end{eqnarray}
for which the particular solution consists of two perpendicular oscillations of the angular momentum vector in the $x$-$y$ plane,
\begin{equation}
    \begin{aligned}
        \hat{L}(t) = & 
        \frac{\omega_\mathrm{h}}{\omega_{\rm p}^2-\omega_\mathrm{h}^2}\frac{\left({{\tau_\mathrm{B}}-\tau_\mathrm{A}}\right)}{2}\cos(\omega_\mathrm{h}t)
        \\
        & \quad +\,\mathrm{i}\left[\frac{\left({{\tau_\mathrm{A}}+\tau_\mathrm{B}}\right)}{2\omega_{\rm p}}\,+ \frac{\omega_{\rm p}}{\omega_{\rm p}^2-
        \omega_\mathrm{h}^2}\frac{\left({\tau_\mathrm{B}}-{\tau_\mathrm{A}}\right)}{2}\sin(\omega_\mathrm{h}t)\right].
    \label{eq:oscillsol}
    \end{aligned}
\end{equation}
As the two oscillations have the same frequencies but different amplitudes, the tip of the angular momentum vector traces an ellipse. As will be seen in Sec.~\ref{sec5}, this dynamic agrees well with the late time evolution of the torus angular momentum observed in numerical simulations.

\subsection{Rotating torque}
\label{rotating}
As a final case, we consider a rotating torque given by $\tau_\mathrm{acc}(t)={\tau_\mathrm{f}}\exp(\mathrm{i}\omega_1 t)$. We obtain the particular solution of the form
\begin{equation}
    \hat{L}(t)=\frac{\mathrm{i}\tau_\mathrm{f}}{\omega_{\rm p} -\omega_1}
    \exp(\mathrm{i}\omega_1 t),
    \label{eq:rotating}
\end{equation}
and the general solution, using Eqs.~(\ref{eq:general}) and (\ref{eq:lt}), reads
\begin{equation}
    L(t) = L_0\exp\left(\mathrm{i}\omega_{\rm p} t\right) + 
    \frac{\mathrm{i}\tau_\mathrm{f}}{\omega_{\rm p} -\omega_1}\exp(\mathrm{i}\omega_1 t).
    \label{eq:rotating-general-solution}
\end{equation}
This solution passes through a resonance at $\omega_1=\omega_{\rm p}$. As $\omega_1\rightarrow\omega_{\rm p}$, the $x$-$y$ component of the angular momentum vector $\boldsymbol{L}$ greatly increases in magnitude, implying that the torus tends to a position in which it ``stands on its edge,'' i.e., the plane of the torus approaches a vertical position (the torus axis becoming horizontal, i.e., tending to line up with the black-hole equatorial plane).  An observer at infinity on the axis of the black hole would then see a constant aspect of the torus (no luminosity variation) as the torus precesses, while a distant observer in the equatorial plane of the black hole would presumably report large luminosity variations, seeing the torus edge-on every half precession period and face-on a quarter of a period later.

\begin{figure*}
    \centering
	\includegraphics[width=\hsize]{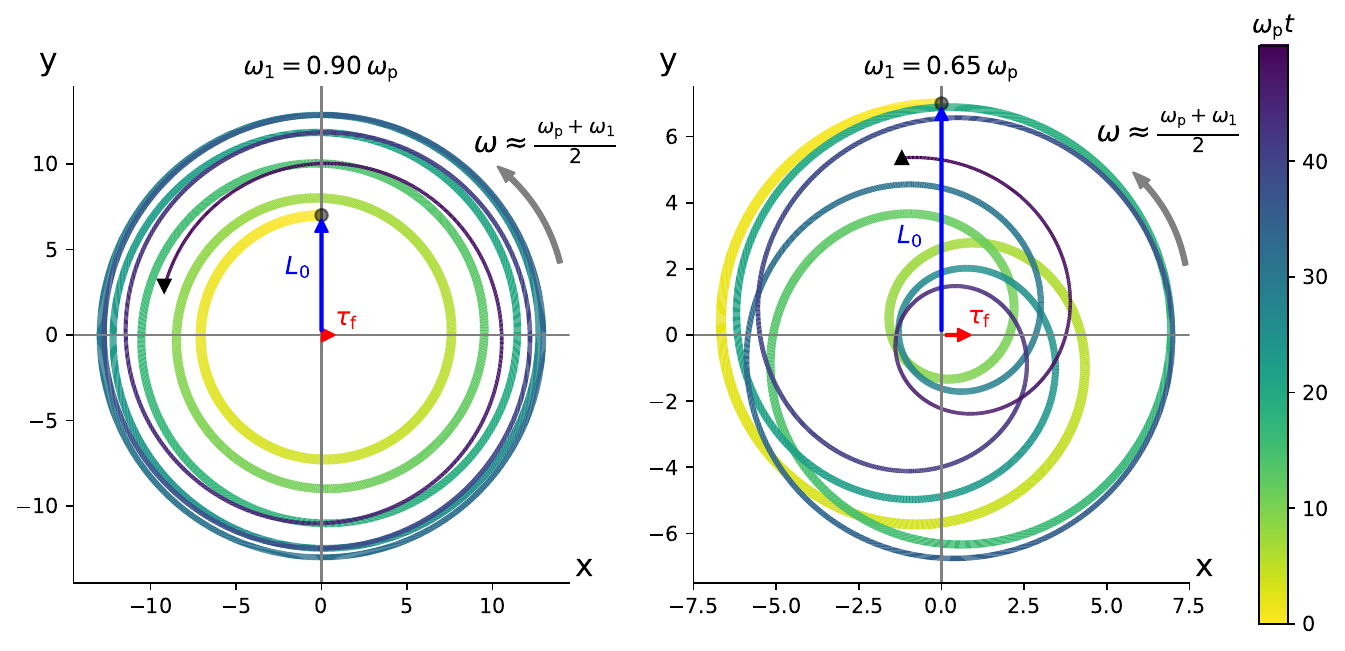}
    \caption{Trajectory of the torus angular momentum vector in the $x$-$y$ plane for an accretion torque rotating with frequency $\omega_1 = 0.9 \omega_\mathrm{p}$ ({\sl left panel}) and $\omega_1 = 0.65\omega_\mathrm{p}$ ({\sl right panel}). In both cases, we use $L(0) = 7{\rm i}$, $\tau_{\rm A} = 1$, and $\omega_\mathrm{p} = 1$. The phase of the trajectory is represented by both the color and the decreasing thickness of the curve, with the starting point at $t=0$ and a point at some later time $t$ marked by a circle and triangle, respectively (for visualization purposes). The torus precesses around the black hole axis with a frequency of $(\omega_{\rm p} + \omega_1)/2$. Additionally, the tilt angle between the black hole and torus oscillates with frequency $(\omega_{\rm p} - \omega_1)$.}
    \label{fig:rotating-trajectory}
\end{figure*}

Perhaps the most interesting situation occurs when the two frequencies are close, but not exactly equal, and the amplitudes of both exponentials in Eq.~\ref{eq:rotating-general-solution} are comparable, i.e.: $\omega_1 \approx \omega_{\rm p}$, but $\omega_1\neq \omega_{\rm p}$, and $|L_0|=|\tau_\mathrm{f}|/(\omega_{\rm p}-\omega_1)$. Then the general solution of Eq.~(\ref{eq:general}), with Eqs.~(\ref{eq:lt}) and (\ref{eq:rotating}), would exhibit beats in which the amplitude of the $x$-$y$ component of the angular momentum vector is modulated at the slow frequency $(\omega_{\rm p}-\omega_1)/2$, while precessing at the fast frequency of $(\omega_1+\omega_{\rm p})/2$. In this case, the modulation means that the midplane of the torus oscillates between a vertical position and the horizontal one, at the frequency of $(\omega_{\rm p}-\omega_1)$. An example is shown in Fig.~\ref{fig:rotating-trajectory}

%-------------------------------------------------------------

\section{Comparison with simulations}
\label{sec5}
The Lense-Thirring precession of the inner torus is a widely accepted model for explaining type-C low-frequency QPOs typically observed in the hard or hard-intermediate states of LMXBs. Often, measurements of the type-C QPO frequency are used to constrain the torus size and the black hole parameters, disregarding the outer thin disk and assuming the torus precesses at the free Lense-Thirring frequency. However, simulations by \citet{Bolimpalli+2023} have demonstrated that the outer thin disk can significantly impact the torus precession rate. Motivated by these findings, we analytically explored the effect of an accretion torque in this paper using simplified forms that mimic various astrophysical scenarios. We now validate our model using the data from simulation a9b15L4, originally detailed in \citet{Bolimpalli+2023}, which is extended to twice the duration initially reported in that paper. The initial setup of this simulation involves a torus extending up to $15\, GM/c^2$ surrounded by a thin slab of gas, both initially misaligned with the black hole spin axis by $15^\circ$. Further details of this simulation can be found in \citet{Bolimpalli+2023, Bollimpalli+2024}. 

\begin{figure*}
	\centering
    \includegraphics[width=\hsize]{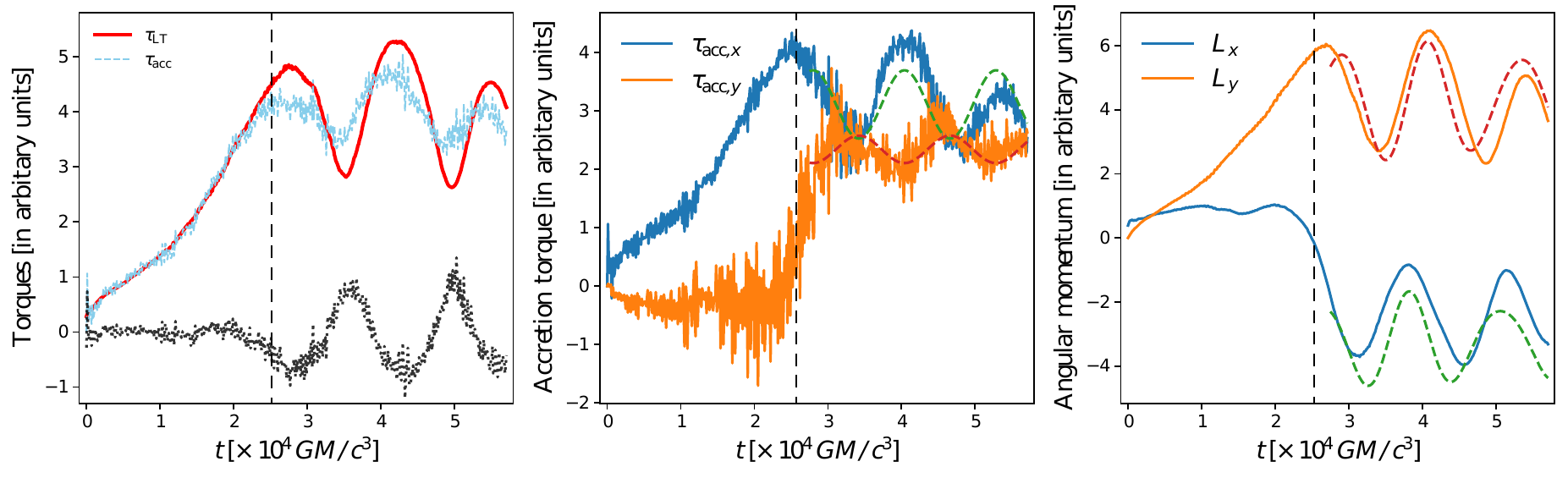}
    \caption{Comparison of our model with the simulation data of \citet{Bollimpalli+2024}. In the {\sl left panel}, we show the magnitude of the accretion (dashed, fluctuating curve) and Lense-Thirring (smooth, solid curve) torques integrated over the torus region, i.e. $5-15\, GM/c^2$. The difference between the two magnitudes (black, dotted curve) remains close to zero in the initial phase of the simulation, up to $\sim\,25000\, GM/c^3$. The {\sl middle panel} shows the $x$ and $y$ components of the net accretion torque on the torus (solid curves), and their respective fits using Eq.~\ref{eq:osilltorq} (dashed curves). Finally, the {\sl right panel} shows the the $x$ and $y$ components of the total angular momentum of the torus (solid curves), and their respective fits using Eq.~\ref{eq:oscillsol} (dashed curves). The vertical, black, dashed lines in all the panels indicate the stop time of the original simulation.}
    \label{fig:sim_cal}
\end{figure*}

For the initial phase of the simulation (up to $\sim\,25000\, GM/c^3$), as originally reported in \citet{Bolimpalli+2023}, the accretion torque and the Lense-Thirring torque are comparable in magnitude, as shown in the first panel of Fig.~\ref{fig:sim_cal}. For these calculations, the accretion torque is evaluated from the angular momentum accretion rate at the outer edge of the torus ($r=15GM/c^2$), while the Lense-Thirring torque is a volume integral of $\boldsymbol{\omega}_\mathrm{p}(\boldsymbol{r})\times\boldsymbol{L}(\boldsymbol{r})$ over the entire volume of the torus ($5-15\,GM/c^2$).
 
As the simulation proceeds past the end point of the original run and into
the extended run time presented here, the torus and the accretion torques begin to satisfy Eq.~\eqref{eq:inho} of our model (a detailed discussion of the precession in the simulation is presented in the Appendix). During this phase (i.e., $t\gtrsim\,25000\, GM/c^3$), the $x$ and $y$ components of the accretion torque exhibit oscillatory behavior, as shown by the solid curves in the second panel of Fig.~\ref{fig:sim_cal}. The oscillatory nature of the torque likely stems from the differential precession of the transition region between the outer thin disk and the torus as the simulation evolves.  We employ the oscillating torque model described in section~\ref{oscillating} to fit the accretion torque exerted on the torus region ($5-15\, GM/c^2$). The best-fit results, obtained using Eq.~\ref{eq:osilltorq}, are shown in the second panel of Fig.~\ref{fig:sim_cal}. {The best fit yielded a constant $\omega_{\rm h} \approx 0.0005$, while $\omega_{\rm p}$ is taken to be the angular-momentum-weighted average of the Lense-Thirring precession frequency for a rigid body (see equation 15 of \citealp{IM2019}):
\begin{equation}
    \bar\omega_p = \frac{\int_{5}^{15}\,\Omega_{\rm LT}\, 
    \Omega\, \Sigma(r)\,  r^3\,{\rm d} r}{\int_{5}^{15}  \Omega\, \Sigma(r)\,  r^3\,{\rm d} r}\,,
    \label{eq:omegap}
\end{equation}
where the local precession frequency is given by the difference between the orbital and vertical epicyclic frequencies, $\Omega_{\rm LT} =  \Omega - \omega_{\perp}$, i.e. 
\begin{equation}
    \Omega_{\rm LT}(r) = \frac{M^{1/2}}{r^{3/2}+a M^{1/2}}
    \left(1-\sqrt{1-\frac{4\,a M^{1/2}}{r^{3/2}} +
    \frac{3a^2}{r^{2}}}\right)   
\end{equation}
is the Lense-Thirring precession frequency of a test particle and $a = 0.9M$ and $M$ are the spin and mass of the black hole, respectively. We directly infer the radial profile of the time-averaged surface density, $\Sigma$, from the simulation. Subsequently, we utilize the best-fit values for $\tau_{\rm A}$ and $\tau_{\rm B}$ from Eq.~\ref{eq:oscillsol} to compute  the $x$ and $y$ components of the total angular momentum $\boldsymbol{L}$, which are shown in the third panel of Fig.~\ref{fig:sim_cal}. }

This exercise underscores the significant influence of the outer thin disk on both the precession rate and the axis of precession. It appears that the initially reported phase of precession in \citet{Bolimpalli+2023} was transient. The later oscillatory behavior highlighted in this section could also be transient; bending waves excited in the inner region propagate outward, inducing precession in the outer thin disk for a brief period before dissipating as they diffuse away. However, since the angular momentum from this region will eventually reach the inner torus, it must impact the precession. To comprehend these intricate dynamics better, our future plan includes running more simulations and for longer periods.

%-------------------------------------------------------------

\section{Discussion and conclusions}
\label{sec:discus}
The two-component model, where an inner hot, geometrically thick accretion flow forms a torus-like structure surrounded by a cold standard accretion disk, is a popular framework for explaining a rich phenomenology of observational properties of accreting black holes in low-mass X-ray binaries. For instance, the photon spectrum may be fitted by a thin accretion disk truncated at some radius and a hotter component identified with the inner torus. When it comes to the observed X-ray variability, the highest frequency QPOs may be explained by the $m=0$ vertical epicyclic mode of the torus \citep{Bursa+2004}, while the type-C QPO may be identified with the $m=-1$ vertical epicyclic mode \citep{Blaes+2006, Fragile16}. Indeed, as demonstrated by numerous numerical simulations, the free torus responds to the Lense-Thirring torque exerted by a black hole by precessing as a rigid body, as long as the sound crossing time remains small compared to the precession period \citep{Fragile+2007, Liska+2018}. However, in reality the two components interact, so that free precession does not correctly describe the motion of the torus. 
 
 In a recent simulation where a torus was placed at the center of a truncated disk, its precession acquired a seemingly complicated character. When considered to be moving about the black hole axis, the torus started precessing at the Lense-Thirring frequency, but subsequently slowed down its precession and eventually seemed to stall. The analytic model we present here does not seem to be able to explain this portion of the simulation. One guess as to why is that the simulation needed longer to adjust from its very artificial initial conditions. 
 
 Subsequently, however, the axis of symmetry of the torus began executing small loops about the position at which it had stalled. We believe this behavior can be understood with the simple rigid-body mechanical model given in Eq.~\ref{eq:inho}, when the angular momentum accreted with the matter from the outer cold disk is taken into account. This additional angular momentum flux corresponds to a new source term (an accretion torque) in the angular momentum conservation equation. In this paper, we have examined the precession of the inner torus in a few of the most natural situations corresponding to different prescribed accretion torques.
 The solutions, presented above, can be described as a precession around an axis which itself executes a definite motion. While, in general, the precession rate of the torus is itself quite robust and remains unaffected by the accretion torque, the accretion torque has a huge impact on the precession amplitude, as well as on the precession axis, or the mean orientation (tilt angle time averaged over the precession period) of the torus.

Already the simplest case of a constant (steady) accretion torque described in Sec.~\ref{steady} brings interesting results. In that case, the inner torus still precesses with the same rate as given by the Lense-Thirring frequency, however, the mean direction of the angular momentum vector corresponds neither to the black-hole spin, nor to the orientation of the outer disk. Instead, it results from the balance between the Lense-Thirring torque $\tau_\mathrm{p}$ and the accretion torque $\tau_\mathrm{f}$ applied by the outer disk, and the angle the mean direction of the torus axis makes with the black-hole spin is $\beta(t) = \arctan[|\hat L|/L_z(t)]$, where $\hat L$ is the (constant) projection onto the $x$-$y$ plane of the angular momentum vector, as determined by the accretion torque, c.f. Eq.~\ref{eq:steady}.

We also considered (in Sec.~\ref{rise}, \ref{healing}, and \ref{oscillating}) three situations in which the accretion torque changes gradually in time. In all cases, a slow ``adiabatic'' change of torque just leads to a sequence of steady precession states, i.e., on top of a free Lense-Thirring precession that remains unaffected, the mean direction of the torus angular momentum gradually changes according to the instant accretion torque. An interesting effect occurs when the rate of torque change becomes comparable to the magnitude of the precession period or larger, i.e. when $(d{\tau}_\mathrm{acc}/dt)\geq\mathcal{O}(\omega_{\rm p})\tau_\mathrm{acc}$. The inner torus responds to these changes by increasing or decreasing the precession amplitude (depending on the relative orientation of the torus angular momentum and the torque change), in addition to the gradual change of the torus mean tilt angle (see Fig.~\ref{fig:healing-trajectory}). As noted already, state transitions in black hole LMXBs are likely accompanied by changes in the overall geometry of the accretion flow, and it is therefore natural to assume that the torque exerted by the outer accretion disk on the inner torus will change, too. Our results show that these changes may even occasionally amplify the Lense-Thirring precession of the inner flow.    

Finally, we have studied situations in which the accretion torque changes periodically in time (Sec.~\ref{rotating}). This may happen when the surrounding parts of the outer disk undergo periodic changes either due to the presence of trapped, low-frequency oscillation modes (e.g. the corrugation mode), or when the accretion rate is modulated periodically. In that case, the torque equation takes the form of a forced linear oscillator. The resonance, occurring when the frequency of the accretion torque matches the precession rate of the inner torus, strongly enhances the inclination of the torus with respect to the black-hole spin. Near the resonance, the torus undergoes low-frequency changes of its inclination, in addition to its precession.

Our findings could have important implications for models of correlations between the spectral properties of black hole LMXBs and the phase of type-C QPOs. Although the precession rate of the inner flow may still be the same Lense-Thirring precession frequency, its radiation may illuminate the outer accretion disk differently than expected, leading to different reflection spectra \citep{Ingram+Done2012b, Ingram+2016} and polarization \citep{Ingram+2015, Fragile25}. This possibility deserves additional study.

%-------------------------------------------------------------

\begin{acknowledgements}
    Part of this work was supported by the German
    \emph{Deut\-sche For\-schungs\-ge\-mein\-schaft, DFG\/} project
    number Ts~17/2--1. Research supported in part by by the Polish NCN grant No. 2019/33/B/ST9/01564. DAB acknowledges support from IIT-Indore, through a Young Faculty Research Seed Grant (project: `INSIGHT'; IITI/YFRSG/2024-25/Phase-VII/02). JH acknowledges support of the the Czech Science Foundation (GA\v{C}R) grant No. 21-06825X. PCF gratefully acknowledges the support of NASA through award No 23-ATP23-0100. The Flatiron Institute is a division of the Simons Foundation. 
\end{acknowledgements}

\bibliographystyle{aa}
\bibliography{lense-thirring.bib}

@ARTICLE{Belloni+2005,
       author = {{Belloni}, T. and {Homan}, J. and {Casella}, P. and {van der Klis}, M. and {Nespoli}, E. and {Lewin}, W.~H.~G. and {Miller}, J.~M. and {M{\'e}ndez}, M.},
        title = "{The evolution of the timing properties of the black-hole transient GX 339-4 during its 2002/2003 outburst}",
      journal = {\aap},
         year = 2005,
        month = sep,
       volume = {440},
       number = {1},
        pages = {207-222},
          doi = {10.1051/0004-6361:20042457},
archivePrefix = {arXiv},
       eprint = {astro-ph/0504577},
 primaryClass = {astro-ph},
       adsurl = {https://ui.adsabs.harvard.edu/abs/2005A&A...440..207B}
}

@ARTICLE{Weng+2021,
       author = {{Weng}, Shan-Shan and {Cai}, Zhen-Yi and {Zhang}, Shuang-Nan and {Zhang}, Wei and {Chen}, Yu-Peng and {Huang}, Yue and {Tao}, Lian},
        title = "{Time-lag Between Disk and Corona Radiation Leads to Hysteresis Effect Observed in Black hole X-Ray Binary MAXI J1348-630}",
      journal = {\apjl},
         year = 2021,
        month = jul,
       volume = {915},
       number = {1},
          eid = {L15},
        pages = {L15},
          doi = {10.3847/2041-8213/ac0a7b},
archivePrefix = {arXiv},
       eprint = {2102.09138},
 primaryClass = {astro-ph.HE},
       adsurl = {https://ui.adsabs.harvard.edu/abs/2021ApJ...915L..15W}
}

@ARTICLE{Shakura+Sunyaev1973,
       author = {{Shakura}, N.~I. and {Sunyaev}, R.~A.},
        title = "{Black holes in binary systems. Observational appearance.}",
      journal = {\aap},
         year = 1973,
        month = jan,
       volume = {24},
        pages = {337-355},
       adsurl = {https://ui.adsabs.harvard.edu/abs/1973A&A....24..337S}
}

@ARTICLE{Haardt+Maraschi1991,
       author = {{Haardt}, F. and {Maraschi}, L.},
        title = "{A Two-Phase Model for the X-Ray Emission from Seyfert Galaxies}",
      journal = {\apjl},
         year = 1991,
        month = oct,
       volume = {380},
        pages = {L51},
          doi = {10.1086/186171},
       adsurl = {https://ui.adsabs.harvard.edu/abs/1991ApJ...380L..51H}
}

@ARTICLE{Haardt+Maraschi1993,
       author = {{Haardt}, Francesco and {Maraschi}, Laura},
        title = "{X-Ray Spectra from Two-Phase Accretion Disks}",
      journal = {\apj},
         year = 1993,
        month = aug,
       volume = {413},
        pages = {507},
          doi = {10.1086/173020},
       adsurl = {https://ui.adsabs.harvard.edu/abs/1993ApJ...413..507H}
}

@ARTICLE{Zdziarski+1998,
       author = {{Zdziarski}, Andrzej A. and {Poutanen}, Juri and {Mikolajewska}, Joanna and {Gierlinski}, Marek and {Ebisawa}, Ken and {Johnson}, W. Neil},
        title = "{Broad-band X-ray/gamma-ray spectra and binary parameters of GX 339-4 and their astrophysical implications}",
      journal = {\mnras},
         year = 1998,
        month = dec,
       volume = {301},
       number = {2},
        pages = {435-450},
          doi = {10.1046/j.1365-8711.1998.02021.x},
archivePrefix = {arXiv},
       eprint = {astro-ph/9807300},
 primaryClass = {astro-ph},
       adsurl = {https://ui.adsabs.harvard.edu/abs/1998MNRAS.301..435Z}
}

@ARTICLE{Zdziarski+2004,
       author = {{Zdziarski}, A.~A. and {Gierli{\'n}ski}, M.},
        title = "{Radiative Processes, Spectral States and Variability of Black-Hole Binaries}",
      journal = {Progress of Theoretical Physics Supplement},
         year = 2004,
        month = jan,
       volume = {155},
        pages = {99-119},
          doi = {10.1143/PTPS.155.99},
archivePrefix = {arXiv},
       eprint = {astro-ph/0403683},
 primaryClass = {astro-ph},
       adsurl = {https://ui.adsabs.harvard.edu/abs/2004PThPS.155...99Z}
}

@ARTICLE{Fabian+2000,
       author = {{Fabian}, A.~C. and {Iwasawa}, K. and {Reynolds}, C.~S. and {Young}, A.~J.},
        title = "{Broad Iron Lines in Active Galactic Nuclei}",
      journal = {\pasp},
         year = 2000,
        month = sep,
       volume = {112},
       number = {775},
        pages = {1145-1161},
          doi = {10.1086/316610},
archivePrefix = {arXiv},
       eprint = {astro-ph/0004366},
 primaryClass = {astro-ph},
       adsurl = {https://ui.adsabs.harvard.edu/abs/2000PASP..112.1145F}
}

@ARTICLE{Krawczynski+Beheshtipour2022,
       author = {{Krawczynski}, H. and {Beheshtipour}, B.},
        title = "{New Constraints on the Spin of the Black Hole Cygnus X-1 and the Physical Properties of its Accretion Disk Corona}",
      journal = {\apj},
         year = 2022,
        month = jul,
       volume = {934},
       number = {1},
          eid = {4},
        pages = {4},
          doi = {10.3847/1538-4357/ac7725},
archivePrefix = {arXiv},
       eprint = {2201.07360},
 primaryClass = {astro-ph.HE},
       adsurl = {https://ui.adsabs.harvard.edu/abs/2022ApJ...934....4K}
}

@ARTICLE{Groselj+2024,
       author = {{Gro{\v{s}}elj}, Daniel and {Hakobyan}, Hayk and {Beloborodov}, Andrei M. and {Sironi}, Lorenzo and {Philippov}, Alexander},
        title = "{Radiative Particle-in-Cell Simulations of Turbulent Comptonization in Magnetized Black-Hole Coronae}",
      journal = {\prl},
         year = 2024,
        month = feb,
       volume = {132},
       number = {8},
          eid = {085202},
        pages = {085202},
          doi = {10.1103/PhysRevLett.132.085202},
archivePrefix = {arXiv},
       eprint = {2301.11327},
 primaryClass = {astro-ph.HE},
       adsurl = {https://ui.adsabs.harvard.edu/abs/2024PhRvL.132h5202G}
}

@ARTICLE{Esin+McClintock+Narayan1997,
       author = {{Esin}, Ann A. and {McClintock}, Jeffrey E. and {Narayan}, Ramesh},
        title = "{Advection-Dominated Accretion and the Spectral States of Black Hole X-Ray Binaries: Application to Nova Muscae 1991}",
      journal = {\apj},
         year = 1997,
        month = nov,
       volume = {489},
       number = {2},
        pages = {865-889},
          doi = {10.1086/304829},
archivePrefix = {arXiv},
       eprint = {astro-ph/9705237},
 primaryClass = {astro-ph},
       adsurl = {https://ui.adsabs.harvard.edu/abs/1997ApJ...489..865E}
}

@ARTICLE{Done+2007,
       author = {{Done}, Chris and {Gierli{\'n}ski}, Marek and {Kubota}, Aya},
        title = "{Modelling the behaviour of accretion flows in X-ray binaries. Everything you always wanted to know about accretion but were afraid to ask}",
      journal = {\aapr},
         year = 2007,
        month = dec,
       volume = {15},
       number = {1},
        pages = {1-66},
          doi = {10.1007/s00159-007-0006-1},
archivePrefix = {arXiv},
       eprint = {0708.0148},
 primaryClass = {astro-ph},
       adsurl = {https://ui.adsabs.harvard.edu/abs/2007A&ARv..15....1D}
}

@ARTICLE{Fragile16,
       author = {{Fragile}, P. Chris and {Straub}, Odele and {Blaes}, Omer},
        title = "{High-frequency and type-C QPOs from oscillating, precessing hot, thick flow}",
      journal = {\mnras},
         year = 2016,
        month = sep,
       volume = {461},
       number = {2},
        pages = {1356-1362},
          doi = {10.1093/mnras/stw1428},
archivePrefix = {arXiv},
       eprint = {1602.08082},
 primaryClass = {astro-ph.HE},
       adsurl = {https://ui.adsabs.harvard.edu/abs/2016MNRAS.461.1356F}
}

@ARTICLE{Fragile25,
       author = {{Fragile}, P. Chris and {Bollimpalli}, Deepika A. and {Schnittman}, Jeremy D. and {Harvey}, Cesare},
        title = "{Polarization Signatures of Quasi-Periodic Oscillations in Simulated Tilted, Truncated Disks}",
      journal = {arXiv e-prints},
         year = 2025,
        month = may,
          eid = {arXiv:2505.11446},
        pages = {arXiv:2505.11446},
          doi = {10.48550/arXiv.2505.11446},
archivePrefix = {arXiv},
       eprint = {2505.11446},
 primaryClass = {astro-ph.HE},
       adsurl = {https://ui.adsabs.harvard.edu/abs/2025arXiv250511446F}
}

@BOOK{Belloni2010,
       author = {{Belloni}, Tomaso},
        title = "{The Jet Paradigm}",
      journal = {Lecture Notes in Physics},
    publisher = {Berlin Springer Verlag},
         year = 2010,
       volume = {794},
          doi = {10.1007/978-3-540-76937-8},
       adsurl = {https://ui.adsabs.harvard.edu/abs/2010LNP...794.....B}
}

@ARTICLE{Singh+2019,
       author = {{Singh}, Chandra B. and {Garofalo}, David and {Kennedy}, Kathryn},
        title = "{The Generalized Hardness-Intensity Diagram for Black Hole and Neutron Star X-Ray Binaries}",
      journal = {\apj},
         year = 2019,
        month = dec,
       volume = {887},
       number = {2},
          eid = {164},
        pages = {164},
          doi = {10.3847/1538-4357/ab4656},
archivePrefix = {arXiv},
       eprint = {1909.08932},
 primaryClass = {astro-ph.HE},
       adsurl = {https://ui.adsabs.harvard.edu/abs/2019ApJ...887..164S}
}

@ARTICLE{Homan+Belloni2005,
       author = {{Homan}, Jeroen and {Belloni}, Tomaso},
        title = "{The Evolution of Black Hole States}",
      journal = {\apss},
         year = 2005,
        month = nov,
       volume = {300},
       number = {1-3},
        pages = {107-117},
          doi = {10.1007/s10509-005-1197-4},
archivePrefix = {arXiv},
       eprint = {astro-ph/0412597},
 primaryClass = {astro-ph},
       adsurl = {https://ui.adsabs.harvard.edu/abs/2005Ap&SS.300..107H}
}

@ARTICLE{Plant+2015,
       author = {{Plant}, D.~S. and {Fender}, R.~P. and {Ponti}, G. and {Mu{\~n}oz-Darias}, T. and {Coriat}, M.},
        title = "{The truncated and evolving inner accretion disc of the black hole GX 339-4}",
      journal = {\aap},
         year = 2015,
        month = jan,
       volume = {573},
          eid = {A120},
        pages = {A120},
          doi = {10.1051/0004-6361/201423925},
archivePrefix = {arXiv},
       eprint = {1309.4781},
 primaryClass = {astro-ph.HE},
       adsurl = {https://ui.adsabs.harvard.edu/abs/2015A&A...573A.120P}
}

@ARTICLE{Kara+2019,
       author = {{Kara}, E. and {Steiner}, J.~F. and {Fabian}, A.~C. and {Cackett}, E.~M. and {Uttley}, P. and {Remillard}, R.~A. and {Gendreau}, K.~C. and {Arzoumanian}, Z. and {Altamirano}, D. and {Eikenberry}, S. and {Enoto}, T. and {Homan}, J. and {Neilsen}, J. and {Stevens}, A.~L.},
        title = "{The corona contracts in a black-hole transient}",
      journal = {\nat},
         year = 2019,
        month = jan,
       volume = {565},
       number = {7738},
        pages = {198-201},
          doi = {10.1038/s41586-018-0803-x},
archivePrefix = {arXiv},
       eprint = {1901.03877},
 primaryClass = {astro-ph.HE},
       adsurl = {https://ui.adsabs.harvard.edu/abs/2019Natur.565..198K}
}

@ARTICLE{Buisson+2019,
       author = {{Buisson}, D.~J.~K. and {Fabian}, A.~C. and {Barret}, D. and {F{\"u}rst}, F. and {Gandhi}, P. and {Garc{\'\i}a}, J.~A. and {Kara}, E. and {Madsen}, K.~K. and {Miller}, J.~M. and {Parker}, M.~L. and {Shaw}, A.~W. and {Tomsick}, J.~A. and {Walton}, D.~J.},
        title = "{MAXI J1820+070 with NuSTAR I. An increase in variability frequency but a stable reflection spectrum: coronal properties and implications for the inner disc in black hole binaries}",
      journal = {\mnras},
         year = 2019,
        month = nov,
       volume = {490},
       number = {1},
        pages = {1350-1362},
          doi = {10.1093/mnras/stz2681},
archivePrefix = {arXiv},
       eprint = {1909.04688},
 primaryClass = {astro-ph.HE},
       adsurl = {https://ui.adsabs.harvard.edu/abs/2019MNRAS.490.1350B}
}

@ARTICLE{DeMarco+2021,
       author = {{De Marco}, B. and {Zdziarski}, A.~A. and {Ponti}, G. and {Migliori}, G. and {Belloni}, T.~M. and {Segovia Otero}, A. and {Dzie{\l}ak}, M.~A. and {Lai}, E.~V.},
        title = "{The inner flow geometry in MAXI J1820+070 during hard and hard-intermediate states}",
      journal = {\aap},
         year = 2021,
        month = oct,
       volume = {654},
          eid = {A14},
        pages = {A14},
          doi = {10.1051/0004-6361/202140567},
archivePrefix = {arXiv},
       eprint = {2102.07811},
 primaryClass = {astro-ph.HE},
       adsurl = {https://ui.adsabs.harvard.edu/abs/2021A&A...654A..14D}
}

@ARTICLE{Rawat+2025,
       author = {{Rawat}, Divya and {M{\'e}ndez}, Mariano and {Garc{\'\i}a}, Federico and {Maggi}, Pierre},
        title = "{Evolution of the Comptonizing medium of the black-hole candidate Swift J1727.8{\textendash}1613 along the hard to hard-intermediate state transition using NICER}",
      journal = {\aap},
         year = 2025,
        month = may,
       volume = {697},
          eid = {A229},
        pages = {A229},
          doi = {10.1051/0004-6361/202453538},
archivePrefix = {arXiv},
       eprint = {2504.06705},
 primaryClass = {astro-ph.HE},
       adsurl = {https://ui.adsabs.harvard.edu/abs/2025A&A...697A.229R}
}

@ARTICLE{Remillard+McClintock2006,
       author = {{Remillard}, Ronald A. and {McClintock}, Jeffrey E.},
        title = "{X-Ray Properties of Black-Hole Binaries}",
      journal = {\araa},
         year = 2006,
        month = sep,
       volume = {44},
       number = {1},
        pages = {49-92},
          doi = {10.1146/annurev.astro.44.051905.092532},
archivePrefix = {arXiv},
       eprint = {astro-ph/0606352},
 primaryClass = {astro-ph},
       adsurl = {https://ui.adsabs.harvard.edu/abs/2006ARA&A..44...49R}
}

@INCOLLECTION{VanDerKlis2006,
       author = {{van der Klis}, M.},
        title = "{Rapid X-ray Variability}",
    booktitle = {Compact stellar X-ray sources},
         year = 2006,
       editor = {{Lewin}, Walter H.~G. and {van der Klis}, Michiel},
       volume = {39},
        pages = {39-112},
    publisher = {Cambridge University Press},
       adsurl = {https://ui.adsabs.harvard.edu/abs/2006csxs.book...39V}
}

@ARTICLE{Psaltis+Belloni+VanDerKlis1999,
       author = {{Psaltis}, Dimitrios and {Belloni}, Tomaso and {van der Klis}, Michiel},
        title = "{Correlations in Quasi-periodic Oscillation and Noise Frequencies among Neutron Star and Black Hole X-Ray Binaries}",
      journal = {\apj},
         year = 1999,
        month = jul,
       volume = {520},
       number = {1},
        pages = {262-270},
          doi = {10.1086/307436},
archivePrefix = {arXiv},
       eprint = {astro-ph/9902130},
 primaryClass = {astro-ph},
       adsurl = {https://ui.adsabs.harvard.edu/abs/1999ApJ...520..262P}
}

@ARTICLE{Wijnands+1999,
       author = {{Wijnands}, Rudy and {van der Klis}, Michiel},
        title = "{The Broadband Power Spectra of X-Ray Binaries}",
      journal = {\apj},
         year = 1999,
        month = apr,
       volume = {514},
       number = {2},
        pages = {939-944},
          doi = {10.1086/306993},
archivePrefix = {arXiv},
       eprint = {astro-ph/9810342},
 primaryClass = {astro-ph},
       adsurl = {https://ui.adsabs.harvard.edu/abs/1999ApJ...514..939W}
}

@ARTICLE{Casella+Belloni+Stella2005,
       author = {{Casella}, P. and {Belloni}, T. and {Stella}, L.},
        title = "{The ABC of Low-Frequency Quasi-periodic Oscillations in Black Hole Candidates: Analogies with Z Sources}",
      journal = {\apj},
         year = 2005,
        month = aug,
       volume = {629},
       number = {1},
        pages = {403-407},
          doi = {10.1086/431174},
archivePrefix = {arXiv},
       eprint = {astro-ph/0504318},
 primaryClass = {astro-ph},
       adsurl = {https://ui.adsabs.harvard.edu/abs/2005ApJ...629..403C}
}

@ARTICLE{Remillard+2002,
       author = {{Remillard}, Ronald A. and {Muno}, Michael P. and {McClintock}, Jeffrey E. and {Orosz}, Jerome A.},
        title = "{Evidence for Harmonic Relationships in the High-Frequency Quasi-periodic Oscillations of XTE J1550-564 and GRO J1655-40}",
      journal = {\apj},
         year = 2002,
        month = dec,
       volume = {580},
       number = {2},
        pages = {1030-1042},
          doi = {10.1086/343791},
archivePrefix = {arXiv},
       eprint = {astro-ph/0202305},
 primaryClass = {astro-ph},
       adsurl = {https://ui.adsabs.harvard.edu/abs/2002ApJ...580.1030R}
}

@ARTICLE{Zdziarski+2023,
       author = {{Zdziarski}, Andrzej A. and {Veledina}, Alexandra and {Szanecki}, Micha{\l} and {Green}, David A. and {Bright}, Joe S. and {Williams}, David R.~A.},
        title = "{Evidence for a Black Hole Spin-Orbit Misalignment in the X-Ray Binary Cyg X-1}",
      journal = {\apjl},
         year = 2023,
        month = jul,
       volume = {951},
       number = {2},
          eid = {L45},
        pages = {L45},
          doi = {10.3847/2041-8213/ace2c9},
archivePrefix = {arXiv},
       eprint = {2304.07553},
 primaryClass = {astro-ph.HE},
       adsurl = {https://ui.adsabs.harvard.edu/abs/2023ApJ...951L..45Z}
}

@ARTICLE{Blandford+Payne1982,
       author = {{Blandford}, R.~D. and {Payne}, D.~G.},
        title = "{Hydromagnetic flows from accretion disks and the production of radio jets.}",
      journal = {\mnras},
         year = 1982,
        month = jun,
       volume = {199},
        pages = {883-903},
          doi = {10.1093/mnras/199.4.883},
       adsurl = {https://ui.adsabs.harvard.edu/abs/1982MNRAS.199..883B}
}

@ARTICLE{Blandford+Znajek1977,
       author = {{Blandford}, R.~D. and {Znajek}, R.~L.},
        title = "{Electromagnetic extraction of energy from Kerr black holes.}",
      journal = {\mnras},
         year = 1977,
        month = may,
       volume = {179},
        pages = {433-456},
          doi = {10.1093/mnras/179.3.433},
       adsurl = {https://ui.adsabs.harvard.edu/abs/1977MNRAS.179..433B}
}

@ARTICLE{Bursa+2004,
       author = {{Bursa}, M. and {Abramowicz}, M.~A. and {Karas}, V. and {Klu{\'z}niak}, W.},
        title = "{The Upper Kilohertz Quasi-periodic Oscillation: A Gravitationally Lensed Vertical Oscillation}",
      journal = {\apjl},
         year = 2004,
        month = dec,
       volume = {617},
       number = {1},
        pages = {L45-L48},
          doi = {10.1086/427167},
archivePrefix = {arXiv},
       eprint = {astro-ph/0406586},
 primaryClass = {astro-ph},
       adsurl = {https://ui.adsabs.harvard.edu/abs/2004ApJ...617L..45B}
}

@ARTICLE{FBG2004,
       author = {{Fender}, R.~P. and {Belloni}, T.~M. and {Gallo}, E.},
        title = "{Towards a unified model for black hole X-ray binary jets}",
      journal = {\mnras},
         year = 2004,
        month = dec,
       volume = {355},
       number = {4},
        pages = {1105-1118},
          doi = {10.1111/j.1365-2966.2004.08384.x},
archivePrefix = {arXiv},
       eprint = {astro-ph/0409360},
 primaryClass = {astro-ph},
       adsurl = {https://ui.adsabs.harvard.edu/abs/2004MNRAS.355.1105F}
}

@ARTICLE{LT1918,
       author = {{Lense}, Josef and {Thirring}, Hans},
        title = "{{\"U}ber den Einflu{\ss} der Eigenrotation der Zentralk{\"o}rper auf die Bewegung der Planeten und Monde nach der Einsteinschen Gravitationstheorie}",
      journal = {Physikalische Zeitschrift},
         year = 1918,
        month = jan,
       volume = {19},
        pages = {156},
       adsurl = {https://ui.adsabs.harvard.edu/abs/1918PhyZ...19..156L}
}

@INCOLLECTION{MR2006,
       author = {{McClintock}, Jeffrey E. and {Remillard}, Ronald A.},
        title = "{Black hole binaries}",
    booktitle = {Compact stellar X-ray sources},
         year = 2006,
       editor = {{Lewin}, Walter H.~G. and {van der Klis}, Michiel},
       volume = {39},
        pages = {157-213},
          doi = {10.48550/arXiv.astro-ph/0306213},
       adsurl = {https://ui.adsabs.harvard.edu/abs/2006csxs.book..157M}
}

@ARTICLE{Bollimpalli+2024,
       author = {{Bollimpalli}, D.~A. and {Fragile}, P.~C. and {Dewberry}, J.~W. and {Klu{\'z}niak}, W.},
        title = "{Truncated, tilted discs as a possible source of Quasi-Periodic Oscillations}",
      journal = {\mnras},
         year = 2024,
        month = feb,
       volume = {528},
       number = {2},
        pages = {1142-1157},
          doi = {10.1093/mnras/stad3975},
archivePrefix = {arXiv},
       eprint = {2312.14876},
 primaryClass = {astro-ph.HE},
       adsurl = {https://ui.adsabs.harvard.edu/abs/2024MNRAS.528.1142B}
}

@ARTICLE{Blaes+2007,
       author = {{Blaes}, Omer M. and {{\v{S}}r{\'a}mkov{\'a}}, Eva and {Abramowicz}, Marek A. and {Klu{\'z}niak}, W{\l}odek and {Torkelsson}, Ulf},
        title = "{Epicyclic Oscillations of Fluid Bodies: Newtonian Nonslender Torus}",
      journal = {\apj},
         year = 2007,
        month = aug,
       volume = {665},
       number = {1},
        pages = {642-653},
          doi = {10.1086/519782},
archivePrefix = {arXiv},
       eprint = {0706.4483},
 primaryClass = {astro-ph},
       adsurl = {https://ui.adsabs.harvard.edu/abs/2007ApJ...665..642B}
}

@ARTICLE{Blaes+2006,
       author = {{Blaes}, O.~M. and {Arras}, P. and {Fragile}, P.~C.},
        title = "{Oscillation modes of relativistic slender tori}",
      journal = {\mnras},
         year = 2006,
        month = jul,
       volume = {369},
       number = {3},
        pages = {1235-1252},
          doi = {10.1111/j.1365-2966.2006.10370.x},
archivePrefix = {arXiv},
       eprint = {astro-ph/0601379},
 primaryClass = {astro-ph},
       adsurl = {https://ui.adsabs.harvard.edu/abs/2006MNRAS.369.1235B}
}

@ARTICLE{Fragile+2007,
       author = {{Fragile}, P. Chris and {Blaes}, Omer M. and {Anninos}, Peter and {Salmonson}, Jay D.},
        title = "{Global General Relativistic Magnetohydrodynamic Simulation of a Tilted Black Hole Accretion Disk}",
      journal = {\apj},
         year = 2007,
        month = oct,
       volume = {668},
       number = {1},
        pages = {417-429},
          doi = {10.1086/521092},
archivePrefix = {arXiv},
       eprint = {0706.4303},
 primaryClass = {astro-ph},
       adsurl = {https://ui.adsabs.harvard.edu/abs/2007ApJ...668..417F}
}

@ARTICLE{Straub+Sramkova2009,
       author = {{Straub}, Odele and {{\v{S}}r{\'a}mkov{\'a}}, Eva},
        title = "{Epicyclic oscillations of non-slender fluid tori around Kerr black holes}",
      journal = {Classical and Quantum Gravity},
         year = 2009,
        month = mar,
       volume = {26},
       number = {5},
          eid = {055011},
        pages = {055011},
          doi = {10.1088/0264-9381/26/5/055011},
archivePrefix = {arXiv},
       eprint = {0901.1635},
 primaryClass = {astro-ph.SR},
       adsurl = {https://ui.adsabs.harvard.edu/abs/2009CQGra..26e5011S}
}

@ARTICLE{Liu+Melia2002,
       author = {{Liu}, Siming and {Melia}, Fulvio},
        title = "{Spin-induced Disk Precession in the Supermassive Black Hole at the Galactic Center}",
      journal = {\apjl},
         year = 2002,
        month = jul,
       volume = {573},
       number = {1},
        pages = {L23-L26},
          doi = {10.1086/341991},
archivePrefix = {arXiv},
       eprint = {astro-ph/0205487},
 primaryClass = {astro-ph},
       adsurl = {https://ui.adsabs.harvard.edu/abs/2002ApJ...573L..23L}
}

@ARTICLE{Ingram+Done+Fragile2009,
       author = {{Ingram}, Adam and {Done}, Chris and {Fragile}, P. Chris},
        title = "{Low-frequency quasi-periodic oscillations spectra and Lense-Thirring precession}",
      journal = {\mnras},
         year = 2009,
        month = jul,
       volume = {397},
       number = {1},
        pages = {L101-L105},
          doi = {10.1111/j.1745-3933.2009.00693.x},
archivePrefix = {arXiv},
       eprint = {0901.1238},
 primaryClass = {astro-ph.SR},
       adsurl = {https://ui.adsabs.harvard.edu/abs/2009MNRAS.397L.101I}
}

@ARTICLE{Papaloizou+Terquem1995,
       author = {{Papaloizou}, John C.~B. and {Terquem}, Caroline},
        title = "{On the dynamics of tilted discs around young stars}",
      journal = {\mnras},
         year = 1995,
        month = jun,
       volume = {274},
       number = {4},
        pages = {987-1001},
          doi = {10.1093/mnras/274.4.987},
       adsurl = {https://ui.adsabs.harvard.edu/abs/1995MNRAS.274..987P}
}

@ARTICLE{Lodado+Facchini2013,
       author = {{Lodato}, Giuseppe and {Facchini}, Stefano},
        title = "{Wave-like warp propagation in circumbinary discs - II. Application to KH 15D}",
      journal = {\mnras},
         year = 2013,
        month = aug,
       volume = {433},
       number = {3},
        pages = {2157-2164},
          doi = {10.1093/mnras/stt878},
archivePrefix = {arXiv},
       eprint = {1306.4333},
 primaryClass = {astro-ph.SR},
       adsurl = {https://ui.adsabs.harvard.edu/abs/2013MNRAS.433.2157L}
}

@INCOLLECTION{Nixon+King2016,
       author = {{Nixon}, Chris and {King}, Andrew},
        title = "{Warp Propagation in Astrophysical Discs}",
    booktitle = {Lecture Notes in Physics},
         year = 2016,
       editor = {{Haardt}, Francesco and {Gorini}, Vittorio and {Moschella}, Ugo and {Treves}, Aldo and {Colpi}, Monica},
       volume = {905},
        pages = {45},
    publisher = {Berlin Springer Verlag},
          doi = {10.1007/978-3-319-19416-5_2},
       adsurl = {https://ui.adsabs.harvard.edu/abs/2016LNP...905...45N}
}

@ARTICLE{Ingram+Done2011,
       author = {{Ingram}, Adam and {Done}, Chris},
        title = "{A physical model for the continuum variability and quasi-periodic oscillation in accreting black holes}",
      journal = {\mnras},
         year = 2011,
        month = aug,
       volume = {415},
       number = {3},
        pages = {2323-2335},
          doi = {10.1111/j.1365-2966.2011.18860.x},
archivePrefix = {arXiv},
       eprint = {1101.2336},
 primaryClass = {astro-ph.SR},
       adsurl = {https://ui.adsabs.harvard.edu/abs/2011MNRAS.415.2323I}
}

@ARTICLE{Ingram+Done2012a,
       author = {{Ingram}, Adam and {Done}, Chris},
        title = "{Modelling variability in black hole binaries: linking simulations to observations}",
      journal = {\mnras},
         year = 2012,
        month = jan,
       volume = {419},
       number = {3},
        pages = {2369-2378},
          doi = {10.1111/j.1365-2966.2011.19885.x},
archivePrefix = {arXiv},
       eprint = {1108.0789},
 primaryClass = {astro-ph.HE},
       adsurl = {https://ui.adsabs.harvard.edu/abs/2012MNRAS.419.2369I}
}

@ARTICLE{Ingram+Done2012b,
       author = {{Ingram}, Adam and {Done}, Chris},
        title = "{The effect of frame dragging on the iron K{\ensuremath{\alpha}} line in X-ray binaries}",
      journal = {\mnras},
         year = 2012,
        month = dec,
       volume = {427},
       number = {2},
        pages = {934-947},
          doi = {10.1111/j.1365-2966.2012.21907.x},
archivePrefix = {arXiv},
       eprint = {1208.0728},
 primaryClass = {astro-ph.HE},
       adsurl = {https://ui.adsabs.harvard.edu/abs/2012MNRAS.427..934I}
}

@ARTICLE{BlackCAT,
       author = {{Corral-Santana}, J.~M. and {Casares}, J. and {Mu{\~n}oz-Darias}, T. and {Bauer}, F.~E. and {Mart{\'\i}nez-Pais}, I.~G. and {Russell}, D.~M.},
        title = "{BlackCAT: A catalogue of stellar-mass black holes in X-ray transients}",
      journal = {\aap},
         year = 2016,
        month = mar,
       volume = {587},
          eid = {A61},
        pages = {A61},
          doi = {10.1051/0004-6361/201527130},
archivePrefix = {arXiv},
       eprint = {1510.08869},
 primaryClass = {astro-ph.HE},
       adsurl = {https://ui.adsabs.harvard.edu/abs/2016A&A...587A..61C}
}

@ARTICLE{SVM99,
       author = {{Stella}, Luigi and {Vietri}, Mario and {Morsink}, Sharon M.},
        title = "{Correlations in the Quasi-periodic Oscillation Frequencies of Low-Mass X-Ray Binaries and the Relativistic Precession Model}",
      journal = {\apjl},
         year = 1999,
        month = oct,
       volume = {524},
       number = {1},
        pages = {L63-L66},
          doi = {10.1086/312291},
archivePrefix = {arXiv},
       eprint = {astro-ph/9907346},
 primaryClass = {astro-ph},
       adsurl = {https://ui.adsabs.harvard.edu/abs/1999ApJ...524L..63S}
}

@ARTICLE{SV98,
       author = {{Stella}, Luigi and {Vietri}, Mario},
        title = "{Lense-Thirring Precession and Quasi-periodic Oscillations in Low-Mass X-Ray Binaries}",
      journal = {\apjl},
         year = 1998,
        month = jan,
       volume = {492},
       number = {1},
        pages = {L59-L62},
          doi = {10.1086/311075},
archivePrefix = {arXiv},
       eprint = {astro-ph/9709085},
 primaryClass = {astro-ph},
       adsurl = {https://ui.adsabs.harvard.edu/abs/1998ApJ...492L..59S}
}

@ARTICLE{IM2019,
       author = {{Ingram}, Adam R. and {Motta}, Sara E.},
        title = "{A review of quasi-periodic oscillations from black hole X-ray binaries: Observation and theory}",
      journal = {\nar},
         year = 2019,
        month = sep,
       volume = {85},
          eid = {101524},
        pages = {101524},
          doi = {10.1016/j.newar.2020.101524},
archivePrefix = {arXiv},
       eprint = {2001.08758},
 primaryClass = {astro-ph.HE},
       adsurl = {https://ui.adsabs.harvard.edu/abs/2019NewAR..8501524I}
}

@ARTICLE{Motta2015,
       author = {{Motta}, S.~E. and {Casella}, P. and {Henze}, M. and {Mu{\~n}oz-Darias}, T. and {Sanna}, A. and {Fender}, R. and {Belloni}, T.},
        title = "{Geometrical constraints on the origin of timing signals from black holes}",
      journal = {\mnras},
         year = 2015,
        month = feb,
       volume = {447},
       number = {2},
        pages = {2059-2072},
          doi = {10.1093/mnras/stu2579},
archivePrefix = {arXiv},
       eprint = {1404.7293},
 primaryClass = {astro-ph.HE},
       adsurl = {https://ui.adsabs.harvard.edu/abs/2015MNRAS.447.2059M}
}

@ARTICLE{eijnden2017,
       author = {{van den Eijnden}, J. and {Ingram}, A. and {Uttley}, P. and {Motta}, S.~E. and {Belloni}, T.~M. and {Gardenier}, D.~W.},
        title = "{Inclination dependence of QPO phase lags in black hole X-ray binaries}",
      journal = {\mnras},
         year = 2017,
        month = jan,
       volume = {464},
       number = {3},
        pages = {2643-2659},
          doi = {10.1093/mnras/stw2634},
archivePrefix = {arXiv},
       eprint = {1610.03469},
 primaryClass = {astro-ph.HE},
       adsurl = {https://ui.adsabs.harvard.edu/abs/2017MNRAS.464.2643V}
}

@ARTICLE{DeFalco+Motta2018,
       author = {{De Falco}, Vittorio and {Motta}, Sara},
        title = "{Polynomial approximation of the Lense-Thirring rigid precession frequency}",
      journal = {\mnras},
         year = 2018,
        month = may,
       volume = {476},
       number = {2},
        pages = {2040-2044},
          doi = {10.1093/mnras/sty361},
archivePrefix = {arXiv},
       eprint = {1802.02866},
 primaryClass = {astro-ph.HE},
       adsurl = {https://ui.adsabs.harvard.edu/abs/2018MNRAS.476.2040D}
}

@ARTICLE{Fragile+Anninos2005,
       author = {{Fragile}, P. Chris and {Anninos}, Peter},
        title = "{Hydrodynamic Simulations of Tilted Thick-Disk Accretion onto a Kerr Black Hole}",
      journal = {\apj},
         year = 2005,
        month = apr,
       volume = {623},
       number = {1},
        pages = {347-361},
          doi = {10.1086/428433},
archivePrefix = {arXiv},
       eprint = {astro-ph/0403356},
 primaryClass = {astro-ph},
       adsurl = {https://ui.adsabs.harvard.edu/abs/2005ApJ...623..347F}
}

@ARTICLE{Teixeira+2014,
       author = {{Morales Teixeira}, Danilo and {Fragile}, P. Chris and {Zhuravlev}, Viacheslav V. and {Ivanov}, Pavel B.},
        title = "{Conservative GRMHD Simulations of Moderately Thin, Tilted Accretion Disks}",
      journal = {\apj},
         year = 2014,
        month = dec,
       volume = {796},
       number = {2},
          eid = {103},
        pages = {103},
          doi = {10.1088/0004-637X/796/2/103},
archivePrefix = {arXiv},
       eprint = {1406.5514},
 primaryClass = {astro-ph.HE},
       adsurl = {https://ui.adsabs.harvard.edu/abs/2014ApJ...796..103M}
}

@ARTICLE{Liska+2018,
       author = {{Liska}, M. and {Hesp}, C. and {Tchekhovskoy}, A. and {Ingram}, A. and {van der Klis}, M. and {Markoff}, S.},
        title = "{Formation of precessing jets by tilted black hole discs in 3D general relativistic MHD simulations}",
      journal = {\mnras},
         year = 2018,
        month = feb,
       volume = {474},
       number = {1},
        pages = {L81-L85},
          doi = {10.1093/mnrasl/slx174},
archivePrefix = {arXiv},
       eprint = {1707.06619},
 primaryClass = {astro-ph.HE},
       adsurl = {https://ui.adsabs.harvard.edu/abs/2018MNRAS.474L..81L}
}

@ARTICLE{Ingram+2016,
       author = {{Ingram}, Adam and {van der Klis}, Michiel and {Middleton}, Matthew and {Done}, Chris and {Altamirano}, Diego and {Heil}, Lucy and {Uttley}, Phil and {Axelsson}, Magnus},
        title = "{A quasi-periodic modulation of the iron line centroid energy in the black hole binary H1743-322}",
      journal = {\mnras},
         year = 2016,
        month = sep,
       volume = {461},
       number = {2},
        pages = {1967-1980},
          doi = {10.1093/mnras/stw1245},
archivePrefix = {arXiv},
       eprint = {1607.02866},
 primaryClass = {astro-ph.HE},
       adsurl = {https://ui.adsabs.harvard.edu/abs/2016MNRAS.461.1967I}
}

@ARTICLE{Bolimpalli+2023,
       author = {{Bollimpalli}, D.~A. and {Fragile}, P.~C. and {Klu{\'z}niak}, W.},
        title = "{Effect of geometrically thin discs on precessing, thick flows: relevance to type-C QPOs}",
      journal = {\mnras},
         year = 2023,
        month = mar,
       volume = {520},
       number = {1},
        pages = {L79-L84},
          doi = {10.1093/mnrasl/slac155},
archivePrefix = {arXiv},
       eprint = {2210.02944},
 primaryClass = {astro-ph.HE},
       adsurl = {https://ui.adsabs.harvard.edu/abs/2023MNRAS.520L..79B}
}

@ARTICLE{Zycki+2016,
       author = {{Zycki}, Piotr T. and {Done}, Chris and {Ingram}, Adam},
        title = "{Energy spectra of X-ray quasi-periodic oscillations in the Lense-Thirring precession model}",
      journal = {arXiv e-prints},
         year = 2016,
        month = oct,
          eid = {arXiv:1610.07871},
        pages = {arXiv:1610.07871},
          doi = {10.48550/arXiv.1610.07871},
archivePrefix = {arXiv},
       eprint = {1610.07871},
 primaryClass = {astro-ph.HE},
       adsurl = {https://ui.adsabs.harvard.edu/abs/2016arXiv161007871Z}
}

@ARTICLE{Veledina+Poutanen2015,
       author = {{Veledina}, Alexandra and {Poutanen}, Juri},
        title = "{Reprocessing model for the optical quasi-periodic oscillations in black hole binaries}",
      journal = {\mnras},
         year = 2015,
        month = mar,
       volume = {448},
       number = {1},
        pages = {939-945},
          doi = {10.1093/mnras/stu2737},
archivePrefix = {arXiv},
       eprint = {1409.6523},
 primaryClass = {astro-ph.HE},
       adsurl = {https://ui.adsabs.harvard.edu/abs/2015MNRAS.448..939V}
}

@ARTICLE{Ingram+2015,
       author = {{Ingram}, Adam and {Maccarone}, Thomas J. and {Poutanen}, Juri and {Krawczynski}, Henric},
        title = "{Polarization Modulation from Lense-Thirring Precession in X-Ray Binaries}",
      journal = {\apj},
         year = 2015,
        month = jul,
       volume = {807},
       number = {1},
          eid = {53},
        pages = {53},
          doi = {10.1088/0004-637X/807/1/53},
archivePrefix = {arXiv},
       eprint = {1505.00015},
 primaryClass = {astro-ph.HE},
       adsurl = {https://ui.adsabs.harvard.edu/abs/2015ApJ...807...53I}
}

@ARTICLE{Bardeen+Patterson1975,
       author = {{Bardeen}, James M. and {Petterson}, Jacobus A.},
        title = "{The Lense-Thirring Effect and Accretion Disks around Kerr Black Holes}",
      journal = {\apjl},
         year = 1975,
        month = jan,
       volume = {195},
        pages = {L65},
          doi = {10.1086/181711},
       adsurl = {https://ui.adsabs.harvard.edu/abs/1975ApJ...195L..65B}
}

@ARTICLE{Wilkins1972,
       author = {{Wilkins}, Daniel C.},
        title = "{Bound Geodesics in the Kerr Metric}",
      journal = {\prd},
         year = 1972,
        month = feb,
       volume = {5},
       number = {4},
        pages = {814-822},
          doi = {10.1103/PhysRevD.5.814},
       adsurl = {https://ui.adsabs.harvard.edu/abs/1972PhRvD...5..814W}
}

@ARTICLE{Kato1989,
       author = {{Kato}, Shoji},
        title = "{Low-frequency, one-armed corrugation waves in relativistic accretion disks.}",
      journal = {\pasj},
         year = 1989,
        month = jan,
       volume = {41},
        pages = {745-761},
       adsurl = {https://ui.adsabs.harvard.edu/abs/1989PASJ...41..745K}
}

@ARTICLE{Silbergleit2001,
       author = {{Silbergleit}, Alexander S. and {Wagoner}, Robert V. and {Ortega-Rodr{\'\i}guez}, Manuel},
        title = "{Relativistic Diskoseismology. II. Analytical Results for C-modes}",
      journal = {\apj},
         year = 2001,
        month = feb,
       volume = {548},
       number = {1},
        pages = {335-347},
          doi = {10.1086/318659},
archivePrefix = {arXiv},
       eprint = {astro-ph/0004114},
 primaryClass = {astro-ph},
       adsurl = {https://ui.adsabs.harvard.edu/abs/2001ApJ...548..335S}
}

@ARTICLE{Kato1993,
       author = {{Kato}, Shoji},
        title = "{Amplification of One-Armed Corrugation Waves in Geometrically Thin Relativistic Accretion Disks}",
      journal = {\pasj},
         year = 1993,
        month = apr,
       volume = {45},
        pages = {219-231},
       adsurl = {https://ui.adsabs.harvard.edu/abs/1993PASJ...45..219K}
}

@ARTICLE{Ipser1996,
       author = {{Ipser}, James R.},
        title = "{Relativistic Accretion Disks: Low-Frequency Modes and Frame Dragging}",
      journal = {\apj},
         year = 1996,
        month = feb,
       volume = {458},
        pages = {508},
          doi = {10.1086/176832},
       adsurl = {https://ui.adsabs.harvard.edu/abs/1996ApJ...458..508I}
}

%-------------------------------------------------------------

\clearpage
\begin{appendix}
\onecolumn
\section{Precession angle in the simulation}
\begin{figure*}
    \centering
	\includegraphics[width=\hsize]{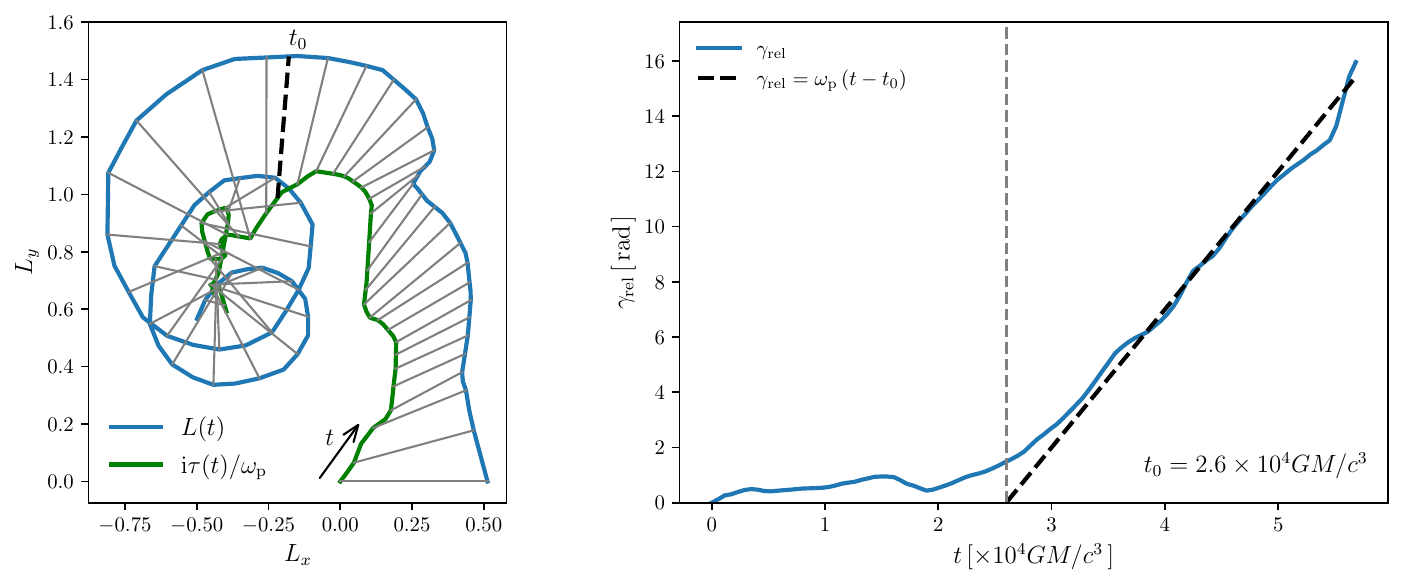}
    \caption{
    {\sl Left:} Evolution of the torus angular-momentum vector $L(t)$ (blue line) and the precession axis given by the torque, $L_\mathrm{axis}(t)=\mathrm{i}\tau_\mathrm{acc}(t)/\omega_\mathrm{p}$ (green line) in the simulation. The gray lines show their relative position at times spaced by constant intervals of $\Delta t=1.148\times10^3 (GM/c^3)$. {\sl Right:} Precession angle relative to the precession axis $L_\mathrm{axis}$ as function of time, $\gamma_\mathrm{rel}(t)=\mathrm{arg}[L(t) - \mathrm{i}\tau(t)/\omega_\mathrm{p}]$. Starting at $t_0=2.6\times 10^3(GM/c^3)$ the torus executes a nearly steady precession with frequency $\omega_\mathrm{p}=5\times10^{-4} c^3/(GM)$.}
    \label{fig:sim_cappend}
\end{figure*}

In Fig.~\ref{fig:sim_cal}, we have used our model to provide fits to the late times in the simulation of \cite{Bollimpalli+2024}. Here, we would like to explain the relevant behavior of the precession angle over the full duration of the simulation. With the knowledge that (the direction of) the precession axis may vary, we are interested in the angle between the angular momentum vector of the torus and the variable precession axis, rather than the fixed spin axis of the black hole. The direction of the precession axis is given by the accretion torque, according to Eq.~\eqref{eq:librium}:
\begin{equation}
   L_\mathrm{axis}(t)=\mathrm{i}\tau_\mathrm{acc}(t)/\omega_\mathrm{p}.
\end{equation}
T his position is tracked in the left panel of Fig.~\ref{fig:sim_cappend} by the green line (beginning next to the time arrow in the figure). The time evolution of the angular momentum vector $L(t)$ during the simulation is tracked by the blue line (the outer edge of the spiral ``staircase''). The gray lines (the ``stairs") join the instantaneous positions of the two vectors, thus, the inclination of the gray lines directly gives the angle between the angular momentum of the torus and the instantaneous precession axis, this is the relative precession angle, $\gamma_\mathrm{rel}$.

As can be seen in the figure, initially $\gamma_\mathrm{rel}$ hardly varies. After about $t=t_0=2.6\times 10^3(GM/c^3)$, the precession angle increases steadily (right panel of Fig.~\ref{fig:sim_cappend}), corresponding to the precession rate of $\omega_\mathrm{p}=5\times10^{-4} c^3/(GM)$. This occurs even while the direction of the precession axis [the tip of the $\mathrm{i}\tau(t)/\omega_\mathrm{p}$ vector] continues to evolve, as can be seen in the left panel.

\end{appendix}

\end{document}